\documentclass[nohyper,notoc]{article} 
\input epsf
\usepackage{bm}
\usepackage{epsfig}
\usepackage{axodraw}
\usepackage{feynmf}
\usepackage{graphicx}
\usepackage{color}
\usepackage{lineno}
\usepackage{cite}

\def\bit{\begin{itemize}}
\def\eit{\end{itemize}}
\def\ben{\begin{enumerate}}
\def\een{\end{enumerate}}
\def\beq{\begin{equation}}
\def\eeq{\end{equation}}
\def\bea{\begin{eqnarray}}
\def\eea{\end{eqnarray}}
\def\bq{\begin{quote}}
\def\eq{\end{quote}}
\def \lsim{\mathrel{\vcenter
     {\hbox{$<$}\nointerlineskip\hbox{$\sim$}}}}
\def \gsim{\mathrel{\vcenter
     {\hbox{$>$}\nointerlineskip\hbox{$\sim$}}}}
\def\gappeq{\mathrel{\rlap {\raise.5ex\hbox{$>$}}
{\lower.5ex\hbox{$\sim$}}}}
\def\lappeq{\mathrel{\rlap{\raise.5ex\hbox{$<$}}
{\lower.5ex\hbox{$\sim$}}}}

\def\ETm{E_T \!  \! \! \! \! \! \! /~~}

\def\hats{\hat{s}}

\def\ETm{E_T \! \! \! \! \! \! /~~}

\def\meg{\mu \to e \gamma}
\def\mec{\mu{\rm -}e ~{\rm conversion}}

\def\topqij{t\to q \ell_i^+ \ell_j^-}

\def\topqme{t\to q \mu^\pm e^\mp}
\def\a{\alpha}
\def\b{\beta}
\def\g{\gamma}

\def\o{\omega}

\def\ifb{fb$^{-1}$}

\begin{document}
\renewcommand{\thefootnote}{\fnsymbol{footnote}}
\begin{flushright}
CERN-PH-TH-2015-145
\end{flushright}
\vskip 1cm
\begin{center}
{\Large {\bf Lepton Flavour Violating top decays at the LHC}}
\vskip 25pt
{\bf   Sacha Davidson $^{2,}$\footnote{E-mail address:
 s.davidson@ipnl.in2p3.fr}, Michelangelo L. Mangano $^{1,}$\footnote{E-mail address:michelangelo.mangano@cern.ch},
 St\'ephane Perries $^{2}$\footnote{E-mail address: s.perries@ipnl.in2p3.fr}
   and Viola Sordini $^{2}$\footnote{E-mail address: v.sordini@ipnl.in2p3.fr}
}
 
\vskip 10pt  
$^1${\it CERN PH-TH, CH-1211 Gen\`eve 23, Switzerland} \\
$^2${\it IPNL, CNRS/IN2P3,  4 rue E. Fermi, 69622 Villeurbanne cedex, France; 
Universit\'e Lyon 1, Villeurbanne;
 Universit\'e de Lyon, F-69622, Lyon, France
}\\
\vskip 20pt
{\bf Abstract}
\end{center}

We consider lepton flavour violating decays of the top quark, mediated
by four-fermion operators.
We compile constraints on  a complete set of 
SU(3)$\times$U(1)-invariant operators, 
arising from their loop contributions
to { rare}  decays and from HERA's single top search.
The  bounds  on $e$-$\mu$ flavour change are more restrictive
than $\ell$-$\tau$; nonetheless
the top could decay to a jet $+ e \bar{\mu}$ with a branching ratio
of order $10^{-3}$. We  estimate that the currently available
LHC data (20~\ifb\ at 8 TeV) could be sensitive to 
$BR(t \to e \bar{\mu}$+ jet) $ \sim 6\times 10^{-5}$,
and extrapolate that   100~\ifb\ at 13 TeV could
reach  a sensitivity of  $ \sim 1 \times 10^{-5}$.

\vskip 20pt  

\setcounter{footnote}{0}
\renewcommand{\thefootnote}{\arabic{footnote}}

\section{Introduction}
\label{intro}

Lepton Flavour Violation(LFV)~\cite{LFV}, meaning 
local interactions that change the flavour of charged
leptons,  should occur because neutrinos have mass and mix.
This motivates sensitive searches for processes
such as $\meg$~\cite{MEG}  and $\mec$~\cite{mec}.
However, the model
responsible for neutrino masses is
unknown,  so it is interesting to 
parametrise LFV with contact
interactions, and to  look for it
everywhere. In this context,  the LHC 
could have the best sensitivity to LFV processes
involving  a heavy leg, such as
the  $Z$~\cite{ZLFVexp,ZLFV,BR}, the Higgs~\cite{BR,hLFVexp,hLFV,FVth}, 
or  a top~\cite{tLFV}.
In this paper, we study
the   LFV top decays $t\to q e^\pm \mu^\mp$,
where $q \in \{u,c\}$.

We  suppose  that these decays are  mediated by a 
four-fermion interaction,
and outline in
Section~\ref{sec:2} the current bounds
on LFV branching ratios of the top.
The  bounds arise from rare decays and
HERA's single top search, and are
discussed in more detail in  the  Appendices.
We find  that, while these bounds place strong 
constraints on some specific Lorentz structures 
for the 4-fermion interactions, they still 
allow for $t \to q e^\pm \mu^\mp$ decays with rates within the LHC reach.
In Section~\ref{sec:experimental_section}, we estimate the
LHC  sensitivity to  
$t\to q e^\pm \mu^\mp$, with  20 fb$^{-1}$
of LHC data at 8 TeV. This estimate relies on simulations of
the background and signal, and is inspired by
the CMS search for $t\to Z q$~\cite{CMS:tZc}. 
The  extrapolation to higher energies
and luminosities is discussed in  Section~\ref{sec:disc}.

{
{\it Quark}-flavour-changing  top decays,
such as $t\to cZ$ and $t\to h c$, 
have been studied in the context of explicit 
models~\cite{FVth,tFVmod,2HDMrev} or described by contact interaction  parametrisations~\cite{tFVCI}, 
and have been 
searched for at the LHC~\cite{CMS:tZc,Aad:2012ij,CMS:2014qxa,Aad:2014dya}.
Quark-flavour-changing (but
lepton-flavour-conserving) three-body decays of the top,
$t \to c f\bar{f}$, where $f$ is a lepton
or quark, have  also  been calculated
in explicit models~\cite{tFV3}.
Top interactions that change
quark and lepton flavour, and, in addition,  baryon and
lepton number,
have  been explored in~\cite{Durieux:2012gj} and searched for by CMS~\cite{Chatrchyan:2013bba}. 
In models
with weak-scale neutrinos  $N$~\cite{etalSoni},
there can be lepton number- and flavour-changing
$W$ decays: 
$W^- \to N \ell \to q \bar{q}' \ell' \ell$,
which could appear in the final state of top decays.
{ In the presence of this decay, $t\bar{t}$  production
could give a final state with three leptons, missing energy and
jets, as in the decay we study (see Figure~\ref{fig:diagram_signal}).
However,  a different combination of leptons and jets 
should reconstruct to the top mass. }
Finally, Fernandez,  Pagliarone,  Ramirez-Zavaleta
and  Toscano~\cite{tLFV} studied almost the same process
as us, $t \to q\tau^\pm \mu^\mp$, but mediated by a (pseudo)-scalar boson.
They obtain  separately the low-energy bounds 
on the quark-and lepton-flavour-changing couplings
of their boson, and obtain that LFV top
branching ratios can be $\sim 10^{-5}$
if the boson mass is $\lsim 2 m_W$.
}


\section{Current bounds }
\label{sec:2}

We are interested in the decays  of a  top 
(or anti-top) to a jet  and a pair of oppositely 
charged leptons of different flavour. In this work,
we focus on the processes $t \to q e^\pm \mu^\mp$,
where $q\in \{u,c\}$, because  the $e$ and
$\mu$ are  easy to identify at the LHC, and
$e \leftrightarrow \mu$ flavour violation
is the most strictly constrained at low energy.
We leave the  decays to $q e^\pm \tau^\mp$ and $q \mu^\pm \tau^\mp$
for a later analysis.

We suppose that these decays are mediated by
 four fermion contact interactions. A complete
list of the $SU(3) \times U(1)$ invariant operators
that we study  is given in Appendix~\ref{ssec:CIs}. 
{We do not impose SU(2) on our operators, because
the scale we will probe is  not far from the
electroweak scale.}
We refer to these LFV operators
as ``top operators''.
Here, as an example,  consider the exchange of a heavy  SU(2)
singlet  leptoquark $S_0$ with couplings
$\lambda_{et} S_0 \overline{e_R} t^c$  and
$\lambda_{\mu c} S_0 \overline{\mu_R} c^c$,
which (after Fierz rearrangement)
  generates the  dimension-6 contact interaction
\beq
\frac{\lambda_{et}^*\lambda_{\mu c}}{2 m^2_{LQ}}
(\overline{\mu}\g^\a P_R e)(\overline{c}\g^\a P_R t)
\equiv -\epsilon^{\mu e ct}_{RR}  \frac{4G_F}{\sqrt{2}}
(\overline{\mu}\g^\a P_R e)(\overline{c}\g^\a P_R t)
= -\epsilon^{\mu e ct}_{RR}  \frac{4G_F}{\sqrt{2}} {\cal O}^{RR}_{\mu e ct}~~~.
\label{LQ}
\eeq
Alternatively, we could define the
operator coefficient as $-1/\Lambda^2$, in which
case $\epsilon \simeq  m_t^2/\Lambda^2$ because
$2 \sqrt{2}G_F \simeq m_t^{-2}$ (we take $m_t = 173.3$ GeV). 
We will quote  low energy bounds on
such interactions as limits on the dimensionless
$\epsilon$s. In the case of our  leptoquark
example, $m_{LQ}\gsim $ 1 TeV to  satisfy
current bounds on second generation
leptoquarks from the LHC~\cite{CMSLQ,Dorsner},
and thus, for $\lambda\lsim 1$, one expects 
$\epsilon^{\mu e ct}_{RR} \lsim 0.02$.

Notice that  
$t \to q e^- \mu^+$ and $t \to q e^+ \mu^-$
are mediated by 
different operators.
Most of the          bounds we
quote will apply to $|\epsilon^{\mu e ct}|^2
+|\epsilon^{e \mu  ct}|^2$, and we will
study the LHC sensitivity assuming
equal rates for 
$t \to q e^- \mu^+$ and $t \to q e^+ \mu^-$.

\subsection{Decay of the top}

In the Standard Model, the top decays
almost always 
to $bW^+$,  at a tree-level rate given by:
\beq
\Gamma (t\to bW) = \frac{g^2|V_{tb}|^2m_t^3}{64\pi m_W^2}
\left(1-\frac{m_W^2}{m_t^2}  \right)^2
\left(1+2\frac{m_W^2}{m_t^2}  \right) \simeq 1.3 ~{\rm GeV}
~~~.
\eeq
In the presence of 
the  operator of Eq.~(\ref{LQ}),
the three-body decay rate 
 is:
\beq
\Gamma (t\to e^+ \mu^-  + c) = 
|\epsilon^{\mu e ct}_{RR}|^2 
 \frac{G_F^2 m_t^5}{192\pi^3} ~~~,
\eeq
so the branching ratio, allowing for all
the operators listed in the Appendix~\ref{ssec:CIs},
 and neglecting  fermion masses other than  the top
(to remove interferences),
 is  
\bea
BR (t \to \ell_i^+ \ell_j^- + q)
\label{BRtop}
&\simeq &\frac{ 1.3}{48\pi^2 } {\Big (}
|\epsilon^{ijqt}_{LL}|^2 + |\epsilon^{ijqt}_{LR}|^2 +
|\epsilon^{ijqt}_{RL}|^2 + |\epsilon^{ijqt}_{RR}|^2
+ \frac{1}{4} {\Big [}|\epsilon^{ijqt}_{S+P,R}|^2 + |\epsilon^{ijqt}_{LQ,R}|^2 -2
{\rm Re}\{\epsilon^{ijqt}_{S+P,R} \epsilon^{ijqt *}_{LQ,R}\} {\Big ]}
 \nonumber\\
&&~~~~ 
+ \frac{1}{4} {\Big [}|\epsilon^{ijqt}_{S+P,L}|^2 + |\epsilon^{ijqt}_{LQ,L}|^2 -
2{\rm Re}\{ \epsilon^{ijqt}_{S+P,L} \epsilon^{ijqt *}_{LQ,L}\} {\Big ]}
+ \frac{1}{4} {\Big [}|\epsilon^{ijqt}_{S-P,L}|^2x + |\epsilon^{ijqt}_{S-P,R}|^2  
{\Big ]} {\Big )}
~~~.
\eea
where $q \in \{ u,c\}$, and  we approximated
$y_t |V_{tb}|\simeq 1$. 
This is small ($\frac{1.3}{48\pi^2 } \simeq  2.8 \times 10^{-3}$), 
due to
three-body phase space.  For the leptoquark example discussed
above, $BR(t \to c e^+ \mu^-) \simeq 10^{-6}$ 
for  $\epsilon^{\mu e ct}_{RR} = 0.02$.

The phase space distribution of the 
$\ell_i^+ \ell_j^- $ and $q$ depends on the Lorentz
and spinor structure of the contact interaction, and could
affect the efficiency of experimental searches for this
decay. The squared matrix-elements for
the individual contact interactions have the form
$|{\cal M}|^2 \propto x(1-x)$, where $x = m_{ab}^2/m_t^2$
and $m_{ab}^2$ is the invariant mass-squared of
a pair of final state fermions $a$ and $b$.
Our study will not take this into account, since we found, 
in some explicit examples, only a small relative 
effect on the selection efficiency (of the order of 5\%).

\subsection{Bounds from flavour physics and HERA}
\label{}

Low energy constraints on four fermion operators involving two leptons
and two quarks have been estimated and  compiled
for  many   operators taken one at a time~\cite{LQ92,Carpentier,Ybook}, 
and  carefully studied for selected flavour combinations
(see {\it e.g.}~\cite{sbtautau}, or global fits
to $b\overline{s} \mu\overline{\mu}$ operators~\cite{BKll}). However, 
even in the more recent compilations~\cite{Carpentier,Ybook},
bounds on LFV operators involving a single top
are not quoted.
In the Appendices \ref{app:meg}
and \ref{app:tloops}, 
we estimate bounds on
such  operators  from their
possible contributions, inside a loop, to rare 
$\mu$,  $B$ and $K$ decays.
In Appendix~\ref{ssec:hera},
we estimate bounds 
from
single top searches at HERA.
Here, we summarise the resulting bounds, and
list in Tables~\ref{tab:topVA} and \ref{tab:topSP}
 the best limits 
on the coefficients of the various operators.
We will find that only the coefficients of some operators
are stringently constrained, while others could mediate LFV top decays within 
the sensitivity of the LHC. 

The current upper limit  $BR(\meg) < 5.7\times 10^{-13}$ \cite{MEG}
severely restricts $e\leftrightarrow \mu$ flavour change. 
For our top operators to contribute, the quark lines
must be closed, which requires at least two loops
and a CKM factor, see the second diagram of 
Figure~\ref{fig:megtopop}. Nonetheless, in the case of scalar
or tensor operators involving $t_R$ this diagram can overcontribute
to $\meg$ by several orders of magnitude, because the
lepton chirality flip is provided by
the operator (rather than $m_\mu$), so the
diagram is enhanced by a factor $m_t/m_\mu$.
We make order-of-magnitude estimates  in Appendix
\ref{app:meg}, and quote the resulting   bounds
in Table~\ref{tab:topSP}.

 Exchanging a $W$ between  the  $t$ and $q$ quark legs
of the top operator will generate an
operator  with  down-type  external
quark legs, see  the left diagram of Figure~\ref{fig:feyn}.
The coefficient of this light quark operator
will be suppressed by a loop, CKM factors, and
various masses. Numerical values for
these suppression factors are given in Table~\ref{tab:nrm} 
of the Appendix; however, their  approximate
magnitude is simple to estimate.
If the top is singlet ($t_R$),  then   
the loop is finite;  in the case of $V \pm A$
interactions, this is because mass
insertions are required on both internal
 quark lines to flip chirality.
In the case of scalar operators, one
internal quark mass for chirality
flip is still required, then terms linear
in the loop momentum vanish, so
the diagram is also proportional to
an external quark mass. For scalar
operators involving $t_L$ (which  require
an  $m_q$ insertion to connect the
$W$ to the $q$ line),  the
best limit can arise from exchanging
a $W$ between the $t$  and a charged lepton leg,
which generates a charged-current operator
as represented in the  second diagram of Figure~\ref{fig:feyn}. 
In the Appendix are also given
current bounds on the coefficients of 
the various light quark operators that the
top operators can induce. Comparing these bounds
to the induced coefficients, gives   the 
limits of Tables~\ref{tab:topVA} and \ref{tab:topSP}
that are  labelled with ($\epsilon$)s.

\begin{figure}[h]
\begin{center}
\epsfig{file=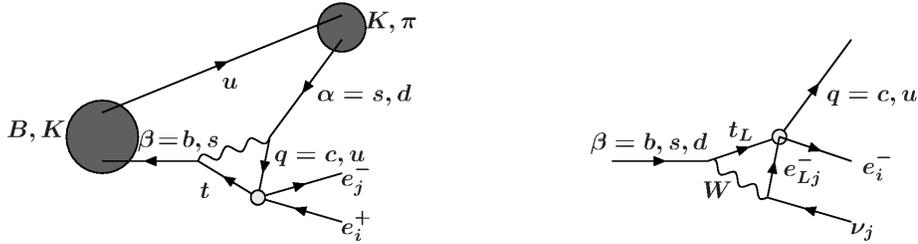, height=4cm,width=14cm}
\end{center}
\caption{Diagrams for generating an 
 LFV operator with light external
quark legs,   by dressing 
a top operator with a $W$ loop.
{  To reduce
index confusion, down-type quarks have
greek indices.}
\label{fig:feyn}}
\end{figure}

\begin{table}[h!]
\begin{center}
\begin{tabular}{|c|c|c|c|c|}
\hline
$ijqt$& ${LL}$ & ${RL}$ & ${LR}$ & ${RR}$\\
&&&& \\
\hline
\hline
$e\mu ut$   &  
$\bm{0.0037}$& 
$\bm{0.0037}$& 
$\bm{0.33}$  & 
$\bm{0.22}$     \\ 
 & 
$(\epsilon^{e\mu ds}_{LL})$ & 
$(\epsilon^{e\mu ds}_{LL})$ & 
(HERA)&
(HERA)
    \\ 
\hline
$e\mu ct$   & 
$\bm{0.015}$&
$\bm{0.015}$  & 1  
 & 1 \\ 
 &   
$(\epsilon^{e\mu ds}_{LL})$ & 
$(\epsilon^{e\mu ds}_{LL})$ & 
$(\epsilon^{e\mu ds}_{LL})$ & 
$(\epsilon^{e\mu ds}_{LL})$
    \\ 
\hline
$e\tau ut$   & 
1.2 & 
1.2 &   
$\sim 1.3$~
& $\sim 0.85~$
\\  &   
$(\epsilon^{e\tau db}_{LL})$ & 
$(\epsilon^{e\tau db}_{LL})$ &
(HERA)&
(HERA)
\\ 
\hline
$e\tau ct$   & 
1 &
1  & 
60  & 60  \\
  &   
$(\epsilon^{ e\tau sb}_{LL})$  
 &
$(\epsilon^{e\tau sb}_{LL})$  
 &  
$(\epsilon^{e\tau sb}_{LL})$ & 
$(\epsilon^{e\tau sb}_{LL})$     \\ 
\hline
$\mu \tau ut$   &   
0.8 & 
0.8 &  $-$ &$-$\\
 &   
$(\epsilon^{ \mu \tau db}_{LL})$ & 
$(\epsilon^{\mu \tau db}_{LL})$  
 &&\\  
\hline
$\mu \tau ct$   & 1.5
 & 1.5
 & 100
 & 100 \\ 
 & 
$(\epsilon^{ \mu \tau sb}_{LL})$ & 
$(\epsilon^{\mu \tau sb}_{LL})$  
 &  
$(\epsilon^{ \mu \tau sb}_{LL})$ & 
$(\epsilon^{\mu \tau sb}_{LL})$  
   \\ 
\hline
\hline
\end{tabular}
\caption{
Constraints on the dimensionless
coefficient  $\epsilon_{XY}^{ijqt}$   of the four-fermion
operator ${\cal O}^{XY}_{ijqt}=$ $2 \sqrt{2}  G_F \left(\bar{e}_{i} \gamma^{\mu} P_{X}
e_{j}\right)\left(\bar{u}_{q}\gamma_{\mu} P_{Y} t\right) $
 (see the
operator list in Appendix
\ref{ssec:CIs}), where $XY$  are  given in the top row,
and  generation indices $ijqt$  are given in the first column.
The bounds {  are on the first
line of each box, their origin
on the second.  They can } arise from the HERA single-top search,
or from the  loop contribution 
to the operator involving lighter fermions
   whose coefficient is
given in parentheses below the bound (see Appendix~\ref{ssec:raredec} for
current experimental bounds on the lighter-fermion operator
coefficients). The $\sim$ bounds are
discussed in Appendix \ref{app:tilde}.
 We expect that $\epsilon < 1$,
so in boldface are the bounds that impose
an upper limit smaller than 1.  
}
\label{tab:topVA}
\end{center}
\end{table}

\begin{table}[h!]
\begin{center}
\begin{tabular}{|c|c|c|c|c|c|c|}
\hline
$ijqt$& $\epsilon_{S - P,R}$  & $\epsilon_{S + P,R}$ & $\epsilon_{T,R}$ 
& $\epsilon_{S-P,L}$ 
& $\epsilon_{S+P,L}$
& $\epsilon_{T,L}$
\\
&&&&&& \\
\hline
\hline
$e\mu ut$   &   
 $\bm{{\cal{O} }( 10^{-2})}$  &  
 $\bm{{\cal{O}} ( 10^{-2})}$    & 
 $\bm{{\cal{O}} ( 10^{-2})}$   & 
$\bm{0.66}$  &
$\bm{ 0.03}$ & \\ 
&$(\meg)$ 
&   $ (\meg)$ 
&$(\meg)$
 & HERA 
&  $ (\epsilon^{e\nu u b}_{CC})$ 
&
 \\ 
\hline
$e\mu ct$   
&$\bm{{\cal{O}} ( 10^{-3})}$   
& $\bm{ {\cal{O}} (10^{-3})}$  
& $\bm{{\cal{O}} ( 10^{-3})}$ 
 &
&     22  
&
 \\ 
&$(\meg)$
& $(\meg)$
&$(\meg)$
& 
&      $ (\epsilon^{e\nu cs}_{CC})$ 
&
  \\ 
\hline
$e\tau ut$   
&23  
& 23  
&
   &
&$\bm{ 0.03}$   
&
 \\ 
&$ (\epsilon^{e\tau db}_{S \pm P,X})$ 
&   $ (\epsilon^{e\tau db}_{S \pm P,X})$ 
&
 &
& $ (\epsilon^{e\nu ub}_{CC})$
&         \\ 
\hline
$e\tau ct$ 
 &   100
 & 100
& 
 &
& 22
& \\ 
&$ (\epsilon^{e\tau db}_{S \pm P,X})$ 
&  $ (\epsilon^{e\tau db}_{S \pm P,X})$ 
&
&      
& $ (\epsilon^{e\nu cs}_{CC})$ 
&
   \\ 
\hline
$\mu \tau ut$ 
 & 21  
& 21    
&
 &
    & $\bm{ 0.03}$
&
 \\ 
&$ (\epsilon^{\mu \tau db}_{S \pm P,X})$ 
&   $ (\epsilon^{\mu \tau db}_{S \pm P,X})$ 
&
 &
& $ (\epsilon^{\mu \nu ub}_{CC})$
&
   \\ 
\hline
$\mu \tau ct$ 
  &  100
 & 100
& 
 &
&   100 
&
 \\ 
&$ (\epsilon^{\mu \tau db}_{S \pm P,X})$ 
&   $ (\epsilon^{\mu \tau db}_{S \pm P,X})$ 
&
 &
&   $ (\epsilon^{\mu \nu cs}_{CC})$
&   \\ 
\hline
\hline
\end{tabular}
\caption{
Constraints on the dimensionless
coefficient  $\epsilon^{ijqt}$, of the
scalar and tensor  four-fermion
interactions.
See Appendix~\ref{ssec:CIs} for operator definitions corresponding to
the subcript of $\epsilon$. The 
generation indices $ijqt$  are given in the first column.
Beneath each bound is given  its origin in parentheses;
$\epsilon^{ij \a \b}_{S \pm P,X}$ are the limits of
Table~\ref{tab:SP}, and 
$\epsilon^{i \nu q \b}_{CC}$ are from
Table~\ref{tab:SPnu}.
{  See the caption of
Table~\ref{tab:topVA} for additional details.}
}
\label{tab:topSP}
\end{center}
\end{table}

HERA collided  protons with positrons (or  electrons)
at a centre of mass energy of 319 GeV,
and searched for single tops in the final state. The
H1 collaboration had a few events with energetic isolated  
leptons and missing energy, consistent with 
  $e^\pm p\to te^\pm + X$ followed by
leptonic decay of the top~\cite{H103,H109}.
However, this signal was not confirmed by the
ZEUS experiment~\cite{ZEUS}, and neither collaboration
had a signal consistent with  hadronic
top decays.  Both collaborations set bounds on
$\sigma (e^\pm p\to e^\pm tX)$; we follow H1,
since they had some events and  a  weaker bound:
\beq
\sigma (e^\pm p\to e^\pm  t  +X) \leq 0.30  ~~ {\rm pb}
 = \frac{2.3\times 10^{-5}}{m_t^2}
~~{\rm  at}~ 95\% CL.
\label{H1bd}
\eeq
Contact interactions of the form
$(\overline{e} \Gamma \mu )
(\overline{u} \Gamma t )$
and
$(\overline{\mu} \Gamma e )
(\overline{u} \Gamma t )$ could respectively mediate
 $e^-p\to \mu^- t X$ and 
 $e^+p\to \mu^+ t  X$. 
As discussed in the Appendix~\ref{ssec:hera}, 
the limit  of Eq.~(\ref{H1bd})
translates to $\epsilon \lsim 0.3 \to 1$ for
the various operators, as given in Tables~\ref{tab:topVA} and \ref{tab:topSP}.

It can be seen from Tables~\ref{tab:topVA} and~\ref{tab:topSP},
that  rare decays give very weak  bounds on some 
 contact interactions of the form
$(\overline{e} \Gamma \tau )
(\overline{u} \Gamma t )$
and
$(\overline{\tau} \Gamma e )
(\overline{u} \Gamma t )$. 
Such  interactions might
have contributed  a signal at HERA via leptonic $\tau$ decays,
so we make some approximate estimates in Appendix \ref{app:tilde},
and include the bounds in the tables with a $\sim$.

\subsection{Implications}
\label{ssec:imp}

{ 
 The current bounds on LFV 
branching ratios of the top can be  obtained from
 Tables~\ref{tab:topVA} and \ref{tab:topSP}, 
and Eq.~(\ref{BRtop}). 
In these tables, the bound on $\epsilon^{ijqt}_{XYZ}$
appears on the first row of each box, where the flavour indices
$ijqt$ are given in the left column, and
the operator label $XYZ$ is given in
the first row (see Appendix~\ref{ssec:CIs} for operator definitions).
In parentheses below the bound, is
a clue to where  it comes from: HERA means
the single top searches at HERA that are
discussed in Appendix~\ref{ssec:hera},
and the ($\epsilon$)s mean the bound
comes from dressing the top operator
with a $W$  loop. For instance, the
bound 0.0037 in the second row and third column of
Table~\ref{tab:topVA}, arises by exchanging
a $W$ between the $t$ and $u$ legs of the
top operator,  which gives the operator
$(\overline{e} \gamma^\rho P_L \mu )
(\overline{d} \gamma_\rho P_Ls )$.
This would mediate the unobserved decay  $K^0 \to e\mu$,
so its coefficient is bounded above, as indicated in
Table~\ref{tab:eedd}. 
(The bounds on lighter-quark operators
relevant to constraining top operators 
are given in Tables~\ref{tab:eedd} -~\ref{tab:SPnu}, and
Table~\ref{tab:nrm} 
gives
 the loop suppression factors with which
the top operators generate the lighter-quark
operators). Translated back to the top operator,
the upper limit on $BR(K^0 \to e\mu)$ gives
the quoted limit on the top operator
coefficient.

In this paper, we are interested in  
top decays to $e^\pm \mu^\mp$, the bounds
on which are given 
in the second and third rows of the tables. 
For many\footnote{The exception is
the $S-P$ operator involving $t_L$, whose loop
suppression factor would involve two light 
quark masses.} of the operators involving the doublet
component of the top ($t_L$; recall that the
last index in the operator label is the top
chirality), the rare decay bounds are restrictive,
implying that these operators  could only induce
$BR(\topqme) \leq 10^{-6}$.
{
Scalar and tensor operators involving $e$, $\mu$ and $t_R$ 
would overcontribute to $\meg$. 
However, there remain
operators which 
are weakly or not constrained,} allowing a  branching
ratio  $\lsim 10^{-3}$. It is therefore interesting
to explore the sensitivity of the LHC to
$ t\to e^\pm \mu^\mp  q$ 
decays.

Finally, it is interesting to consider how large the $\epsilon$ coefficient
of the top operators can be. Some of the upper bounds quoted
in Tables~\ref{tab:topVA} and~\ref{tab:topSP} are $\gg$ 1,
and should not be interpreted as relevant constraints
\footnote{They are given so that in the future, if the
experimental bounds improve, the limits can be
obtained by simply rescaling the number in the tables.
For instance, if the upper bound on $B\to \tau \ell X$
decays were to improve by two orders of magnitude,
the limit on some $\epsilon$s would be divided by 10,
and become marginally relevant.}. Indeed, the
width of the top is given by D0~\cite{largeurt} as
$2.0 \pm 0.5$ GeV (the theoretical decay rate to
$bW$ is 1.3 GeV), which constrains $\epsilon^{ijqt}_{XYZ} < 10$-$20$. 
Furthermore, phenomenological prejudice and
the leptoquark example of Eq.~(\ref{LQ}),  suggest
that $\epsilon<1$, because the
three-body decay should be mediated by 
sufficiently heavy ($m>m_t)$ particles,
with sufficiently small couplings to
have not yet been detected. We therefore
quote in bold face the ``relevant'' bounds that 
impose $\epsilon < 1$.
 }


\section{ $\bm{t\to e^\pm \mu^\mp q}$  at the 8 TeV LHC}
\label{sec:experimental_section}

 In this Section, we estimate the sensitivity of current LHC data to  the LFV 
top decays
$t \to q e^{\pm}\mu^{\mp}$, where $q=u, c$.
 We consider  strong production of a $t\bar{t}$ pair,
because this is the most abundant source of tops at the LHC,
 followed by the leptonic decay on one top, and the LFV decay of the other.
This is illustrated in Figure~\ref{fig:diagram_signal}, and gives a final state
containing three isolated muons or electrons \footnote{The final states where the $W$ decays to $\tau^{-}\bar{\nu_{\tau}}$ are not directly targeted by this search.
The fact that such processes, followed by leptonic $\tau$ decays, can pass our selection is taken into account in the signal efficiency,
as explained in the following.}, which has small Standard Model 
backgrounds. 

\subsection{Simulation setup}
This study is performed for proton-proton collisions at the LHC, with center of mass energy of 8 TeV and an integrated luminosity of 20 $\mathrm{fb}^{-1}$, corresponding to the LHC Run1. The details about the signal and background generation are given in Section~\ref{subsec:gen}. 
The detector simulation is carried out by Delphes~\cite{deFavereau:2013fsa} using a CMS setup parametrization.

Delphes uses particle-flow-like reconstruction.
The relative isolation of leptons is calculated from the total $p_T$ of the particles inside a cone of $\Delta R$ around the lepton direction ($\Delta R=0.3$ for electrons and $0.4$ for muons), divided by the $p_T$ of the lepton. 
Jets are clustered using the fastjet package~\cite{Cacciari:2011ma} with the Anti-kt~\cite{Cacciari:2008gp} algorithm with distance parameter $R=0.5$.
The b-tagging performances are tuned on the typical efficiency and fake rate obtained in CMS. 

For this study, no additional interactions in the same or neighbouring bunch crossing (pileup) are simulated.

\subsection{Signal and SM backgrounds generation}
\label{subsec:gen}
The signal is generated with PYTHIA 8.205~\cite{Sjostrand:2007gs} 
using tune 4C. Top quarks are pair produced,
then one top is  forced to decay to charm, $\mu^\pm$, and $ e^\mp$, 
with equal probability between $\mu^+ \, e^-$ and $\mu^- \, e^+$. 
The decay products are distributed according to the 
available phase space. 100k events have been generated 
both for LFV top and anti-top decays.

\begin{figure}[h]
\begin{center}
\epsfig{file=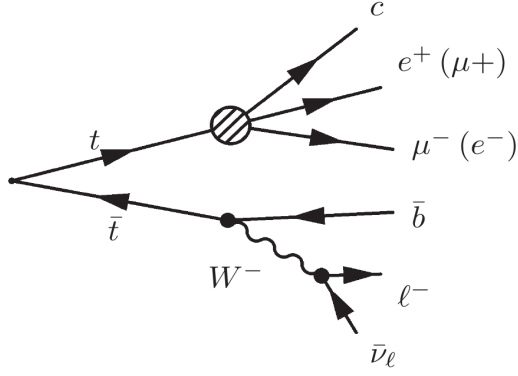,height=5cm,width=9cm}
\end{center}
\caption{Feynman diagram for the considered signal where $\ell=e$ or $\mu$ (the conjugate diagram is also considered).}
    \label{fig:diagram_signal}
\end{figure}

The backgrounds for this search, listed in
Table~\ref{tab:all_bkg_xsec}, are processes that can give rise to
three isolated leptons and at least 2 jets in the final state.  Most
of them are related to the production of real isolated leptons, e.g.
from a top pair or vector bosons in the final state.  In the table,
are also shown the details about the number of generated events and
production cross section for 8 TeV and 13 TeV proton-proton
collisions.  The number of generated events refers to the generation
at 8 TeV.  The $t\bar{t}$ cross section is calculated with the Top++2.0 program to
next-to-next-to-leading order in perturbative QCD, including
soft-gluon resummation to next-to-next-to-leading-log order
(see~\cite{Czakon:2011xx} and references therein), and assuming a
top-quark mass of $m_{t} = 173.3$ GeV.  When an explicit calculation
was not available, the cross sections have been calculated with the
MCFM package~\cite{Campbell:2010ff}, version $7.0$. The kinematic cuts
used for the calculation are also shown in the table.

\begin{table}[h!]
\centering
\begin{tabular}{|c|c|c|c|c|}
\hline
Process & Events & $\sigma$ [pb] (8 TeV) & $\sigma$[pb] (13 TeV) & Source \\
\hline
$t\bar{t}$ (2l2$\nu$2b)                &  12M   & 26.19 & 86.26 & Top++2.0 (NNLO)~\cite{Czakon:2011xx}\\
\hline
WW+jets (2l2$\nu$, $m_{ll}>10 GeV/c^2$)&  1M    & 5.84   & 11.57 & MCFM~\cite{Campbell:1999ah} \\
\hline
ZZ+jets (2l2q, $m_{ll}>10 GeV/c^2$)    &  2.5M  & 2.71   & 5.35 & MCFM~\cite{Campbell:1999ah}\\
\hline
ZZ+jets (2l2$\nu$, $m_{ll}>10 GeV/c^2$)&  1M    & 0.774  & 1.53 & MCFM~\cite{Campbell:1999ah} \\
\hline
ZZ+jets  (4l, $m_{ll}>10 GeV/c^2$)     &  2.5M  & 0.390  & 0.738 & MCFM~\cite{Campbell:1999ah}\\
\hline
WZ+jets (2l2q, $m_{ll}>10 GeV/c^2$)    &  1M    & 2.37   & 4.60 & MCFM~\cite{Campbell:1999ah}\\
\hline
WZ+jets (3l$\nu$, $m_{ll}>10 GeV/c^2$) &  1M    & 1.15   & 2.23 & MCFM~\cite{Campbell:1999ah}\\
\hline
$t\bar{t}$W+jets                       &  1M    & 0.212  & 0.612 & MCFM~\cite{Campbell:1999ah}\\
\hline
$t\bar{t}$Z+jets                       &  1M    & 0.192  & 0.798 & aMC@NLO~\cite{Alwall:2014hca} \\
\hline
tbZ+jets                               &  2M    & 0.014  & 0.047 & aMC@NLO~\cite{Alwall:2014hca} \\
\hline
\end{tabular}
\caption{
Number of events generated, and  cross sections at NLO (except for
$t\bar{t}$+jets),
for each background category. Here $l=e, \mu, \tau$.\label{tab:all_bkg_xsec}}
\end{table}

The leading order (LO) matrix element generator, MADGRAPH 5~\cite{MG5}, with CTEQ6 parton distribution functions, 
is used to generate top pair production, and associated production of 
a top pair and a vector boson ($ttW$, $ttZ$). 
MADGRAPH, interfaced with tauola for $\tau$ decays, is 
used to generate vector-vector production ($WW$, $WZ$ and $ZZ$) 
and the contribution of weak processes giving rise to final
 states with one top quark, one $b$ quark and a $Z$ boson (decaying to leptons). 
For the vector-vector production, we only consider final states 
with at least two real charged leptons.  This means  that for the $WW$ system,
the considered final states 
 are 2 charged leptons and 2 neutrinos; for $WZ$, they  are  
3 charged leptons and one neutrino or 2 charged leptons and 2 quarks; 
and for $ZZ$, they  are 2 charged leptons and 2 neutrinos, 2 charged 
leptons and 2 quarks, or 4 charged leptons. 
In all cases, MADGRAPH accounts for the presence of 
up to two additional jets at matrix-element level, and the 
hadronisation is carried out by PYTHIA 8.205. 
The details about the SM backgrounds simulation and cross sections are shown in 
Table~\ref{tab:all_bkg_xsec}.

\subsection{Event selection}
\label{sec:sel}

The signal for this search is   $t\bar{t}$ production, followed by the lepton flavor violating decay of one top (which will be denoted as LFV top in the following),
and the leptonic decay of the other (standard top, in the following).  
The number of expected signal events is given by:
\begin{eqnarray}
N_{SIG} = \mathcal{L}\cdot \epsilon_{SIG}\cdot2\cdot \sigma_{t\bar{t}} \cdot BR(W\to l\nu) \cdot BR(\topqme) ~~~,
\label{eq:sig_xsec}
\end{eqnarray}
where $l \in \{e,\mu,\tau\}$, $ \mathcal{L}$ the integrated luminosity, $\epsilon_{SIG}$ the selection efficiency on the signal, and $\sigma_{t\bar{t}}=246.7$ pb 
(see~\cite{Czakon:2011xx} and references therein).

The considered signature is  3 isolated leptons (with one pair of opposite sign and opposite flavor from the LFV decay), 2 jets (one of which is a $b$-jet), and missing transverse energy. 
For the event selection, we consider only muons of $p_T>20$ GeV and $|\eta|<2.4$, electrons of $p_T>20$ GeV and $|\eta|<2.5$, and jets of $p_T>30$ GeV and $|\eta|<2.4$. 
These criteria are comparable to those used in real analyses by the CMS or ATLAS collaborations. A muon is considered isolated if its relative isolation value is less that 0.12, and 
an electron is considered isolated if its relative isolation value is less that 0.1. 
We select events containing:
\ben
\item
 exactly 3 isolated charged leptons (electrons or muons), two of which must be of opposite sign and opposite flavor. 
\item
Events are requested to contain at least 2 jets, and 
\item exactly one b-tagged jet, and 
\item the missing transverse energy has to be higher than 20 GeV. 
\item 
In order to exclude events where two of the isolated leptons come from a real Z boson, we reject events containing any pair of opposite sign isolated muons or electrons with invariant mass between 78 and 102 GeV/c$^2$. This cut is particularly helpful in rejecting background arising from $t\bar{t}$ associated production with a Z.

\item
The charged lepton that does not belong to the pair of opposite sign and opposite flavor leptons is assigned to the standard top in the event, and assumed to come from the W decay (bachelor lepton). 
Following a common procedure in reconstruction of $t\bar{t}$ semi-leptonic events, the $x$ and $y$ components of the missing transverse energy are taken as a measurement of the neutrino 
$p_{x}$ and $p_{y}$, and the longitudinal component of the neutrino momentum is calculated imposing that the invariant mass of the system composed of the bachelor lepton and the neutrino must equal the mass of the W boson. 
The bachelor lepton and the neutrino 4-momenta are then combined with that of the b-tagged jet, to build a candidate standard top.
When more choices of the bachelor lepton are possible (there can be up to 2 possible pairs of opposite sign opposite flavor charged leptons in one event), all are considered and the one giving the best standard top mass is chosen.  
We reject events in which the invariant mass of the standard top candidate is more than 45 GeV away from the nominal top mass. 
After the choice of the bachelor lepton, there is only one possible pair of opposite sign and opposite flavor leptons in each event. This is combined with all good (non b-tagged) jets 
present in the event to build a list of candidates for the LFV top. 

 Events are requested to have at least one combination giving a LFV top mass within 25 GeV of the nominal value.

\een
The efficiency of the final selection on signal events is
\beq
 \epsilon_{SIG}= (1.85\pm0.03)\% ~~~,
\label{efficiency}
\eeq
where the uncertainty is statistical only. 
The signal efficiency is calculated on $t\bar{t}$ events where one top 
decays through $\topqme$, and the other one decays to a $b$ quark and a $W$, which subsequently decays to a charged lepton (e, $\mu$ or $\tau$) and a neutrino, and is defined as the fraction of such events passing the selection criteria.  

The number of expected events, for the signal and for each background category, on 20 fb$^{-1}$ of proton-proton data at 8 TeV
is shown in Table~\ref{tab:cut_flow}, for different subsequent selection requirements.

\begin{table}
\centering
\begin{tabular}{|c|c|c|c|c|c|c|c|}
\hline
Process     & no selection & step 1 & step 2  & step 3 & step 4 & step 5 & step 6 \\
\hline
Signal      & 202.57   &  32.98 $\pm$ 0.17 & 22.66 $\pm$ 0.14 & 9.12 $\pm$ 0.09 & 8.20 $\pm$ 0.09 & 7.50 $\pm$ 0.09 & 3.75 $\pm$ 0.06 \\  
\hline
$t\bar{t}$  & 542806   &  14.78 $\pm$ 0.81 & 10.51 $\pm$ 0.69 & 4.36 $\pm$ 0.44 & 4.36 $\pm$ 0.44 & 3.55 $\pm$ 0.40 & 0.63 $\pm$ 0.17 \\
WW+jets     & 116760   &   0.93 $\pm$ 0.33 &  0.35 $\pm$ 0.20 & $< 0.35$        & $< 0.35$        & $< 0.35$        & $< 0.35$        \\
ZZ+jets     &  72900   & 353.74 $\pm$ 0.95 & 82.50 $\pm$ 0.47 & 3.74 $\pm$ 0.10 & 1.60 $\pm$ 0.07 & 0.25 $\pm$ 0.03 & 0.03 $\pm$ 0.01 \\
WZ+jets     &  63360   & 852.21 $\pm$ 4.04 &182.96 $\pm$ 1.90 & 8.70 $\pm$ 0.42 & 7.62 $\pm$ 0.39 & 0.74 $\pm$ 0.12 & 0.04 $\pm$ 0.03 \\
$t\bar{t}$W &   4240   &   9.36 $\pm$ 0.24 &  7.67 $\pm$ 0.22 & 3.59 $\pm$ 0.15 & 3.45 $\pm$ 0.14 & 3.10 $\pm$ 0.14 & 0.27 $\pm$ 0.04 \\
$t\bar{t}$Z &   3840   &  17.25 $\pm$ 0.33 & 16.44 $\pm$ 0.32 & 7.72 $\pm$ 0.22 & 7.16 $\pm$ 0.21 & 1.85 $\pm$ 0.11 & 0.22 $\pm$ 0.04 \\
tbZ         &    282   &   5.75 $\pm$ 0.03 &  3.59 $\pm$ 0.02 & 1.51 $\pm$ 0.01 & 1.37 $\pm$ 0.01 & 0.13 $\pm$ 0.01 & 0.01 $\pm$ 0.01 \\
\hline
\end{tabular}
\caption{Number of expected events for a luminosity of 20 fb$^{-1}$, at various steps of the selection, for the signal process normalized to a branching ratio $BR(t\to qe\mu)=6.3 \cdot 10^{-5}$, and the various backgrounds considered in this study normalized to their NLO cross sections. All uncertainties are statistical only. The considered backgrounds are the same as in Table~\ref{tab:all_bkg_xsec}, grouped in wider categories. In particular, the numbers in the "no selection" column are relative only to the final states detailed in Table~\ref{tab:all_bkg_xsec}. The steps in the selection are, step 1: 3 leptons with two opposite sign opposite flavour, step 2: at least 2 jets, step 3: exactly one b-tag, step 4: missing $E_T$ greater
than 20 GeV, step 5: Z boson veto, step 6: invariant masses cuts (see Section~\ref{sec:sel} and Figure~\ref{fig:top_invMass}). Uncertainties are statistical only. \label{tab:cut_flow}}
\end{table}

\subsection{Results and expected limits on  the branching ratio}
\label{sec:limits}

The selection and its efficiency, on signal and background, are discussed in Section~\ref{sec:sel},
and summarised in Table~\ref{tab:cut_flow}.
Assuming a 
branching ratio of $BR(\topqme)=6.3 \cdot 10^{-5}$  
for the signal,  an uncertainty of 2.5\% on the luminosity,
and 20 fb$^{-1}$ of data,
 we would expect $N_{SIG}=3.75\pm0.06$ 
signal events,  to compare to the $N_{BKG}=1.20\pm 0.18$ expected events from 
known backgrounds.

In Figure~\ref{fig:top_invMass}, we show the invariant mass of the LFV top candidate (left) and the standard top candidate (right), in events passing all the cuts  except those on the masses themselves.

\begin{figure}[htbp!]
\includegraphics[width=0.5\textwidth,angle =0]{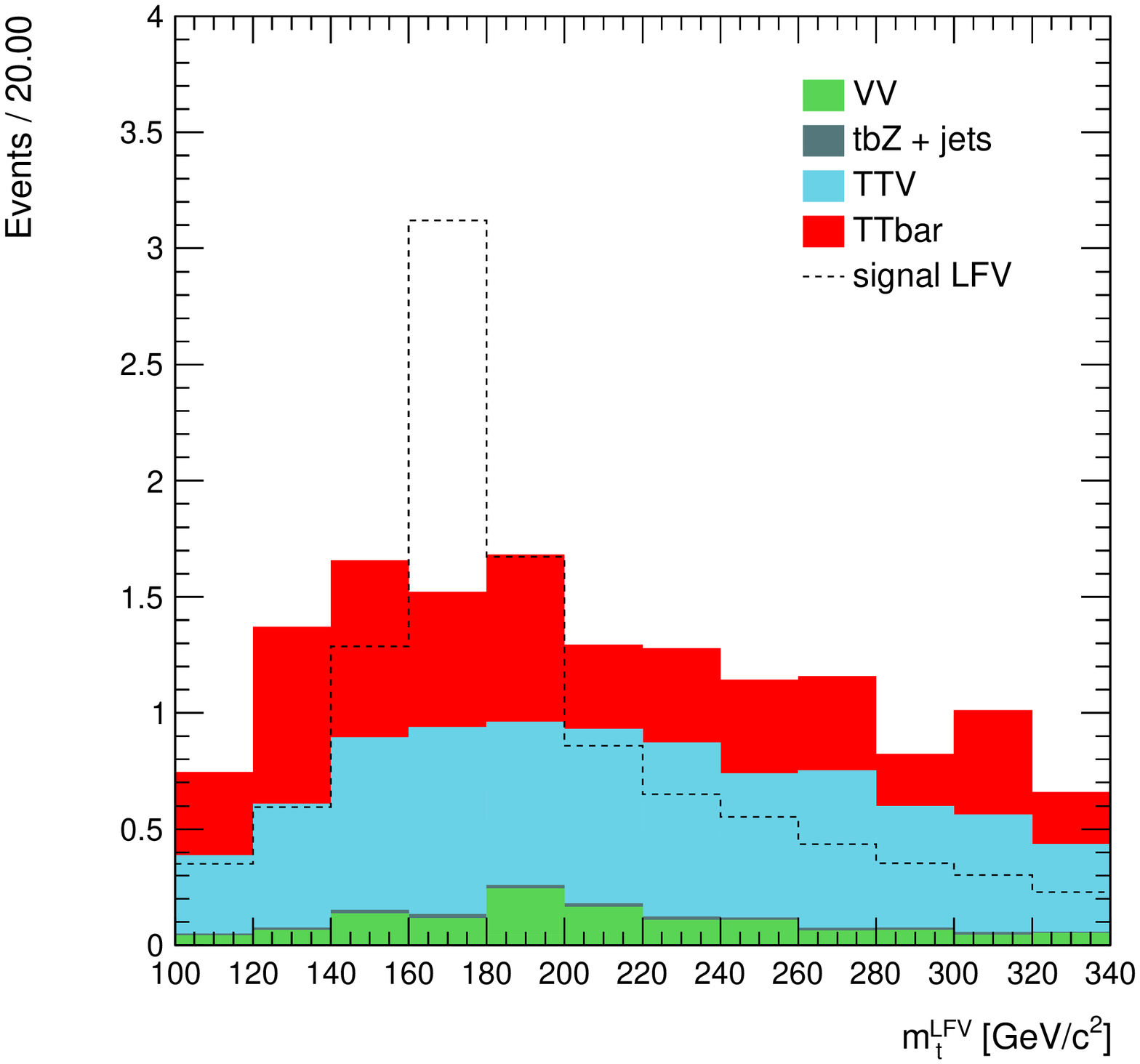}
\includegraphics[width=0.5\textwidth,angle =0]{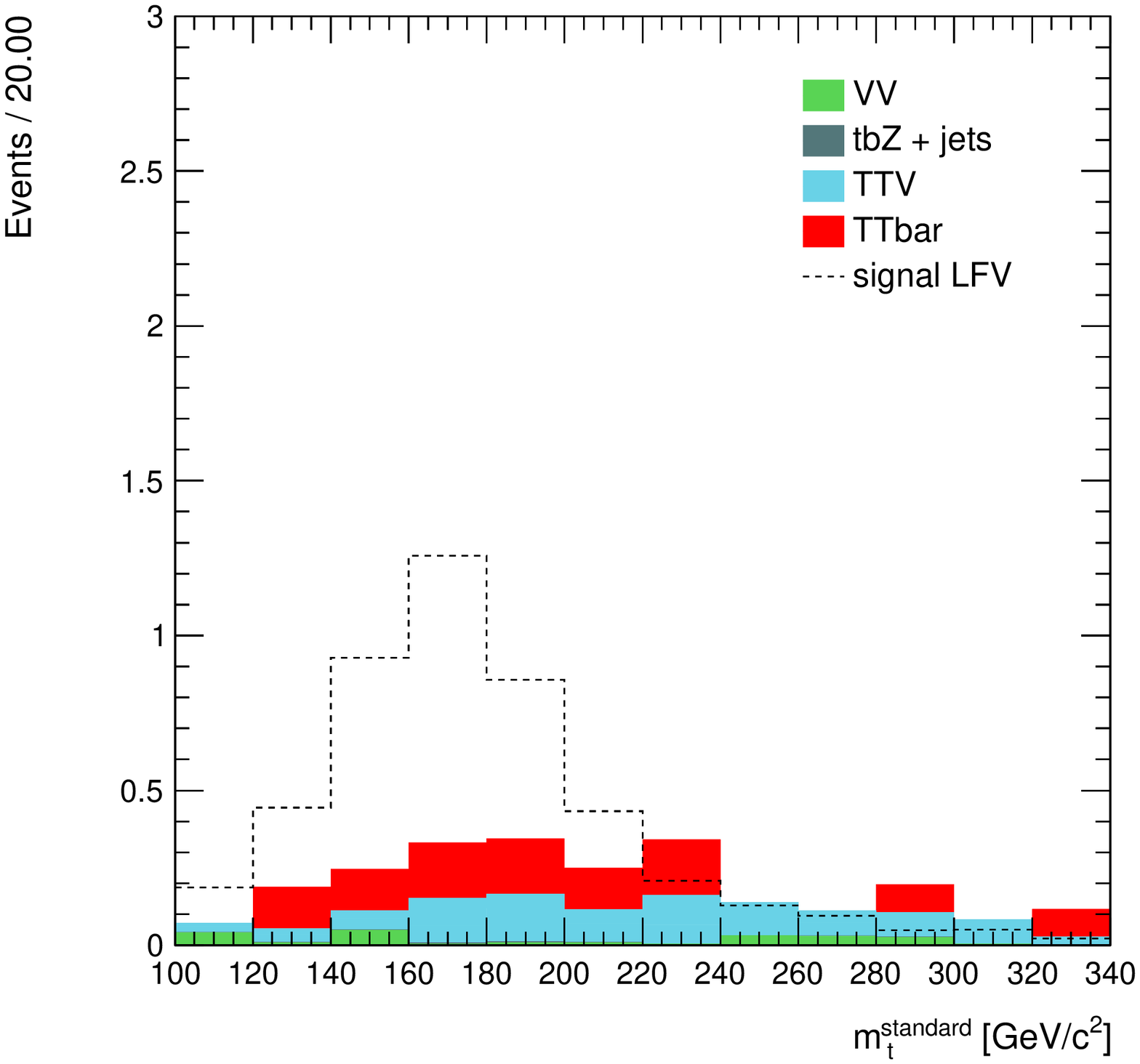}
\caption{Invariant mass of the LFV top candidates (left) and standard top candidate (right) in events passing all the selection, apart from the cut on the masses themselves. The different background contributions are shown in filled histograms and stacked, normalised to the number of events expected in 20 fb$^{-1}$. For comparison, the distribution for signal events is also shown as a dashed line, normalised to the number of events expected in 20 fb$^{-1}$, for an example signal branching ratio $BR(\topqme) = 6.3 \cdot 10^{-5}$, equal to the expected limit extracted in Section~\ref{sec:limits}. 
\label{fig:top_invMass} }
\end{figure}

In order to evaluate the sensitivity of this search for $\topqme$, 
we calculate the expected upper limit that 
could be set, in the case of absence of the signal. 
The calculation is based on the number of expected background 
events surviving the final selection, in 20 fb$^{-1}$ of 8 TeV LHC data, 
so the result can be interpreted as the possible upper limit 
the CMS or ATLAS collaborations (alone) could expect  to set with Run1 data, if the signal were not there.

A 95$\%$ confidence level (CL) upper limit on the branching fraction of $\topqme$ is calculated using the modified frequentist approach 
(CL$_{s}$ method~\cite{Junk:1999kv,Read:2002hq}), as it is implemented in the RooStats framework~\cite{Moneta:2010pm}.
Based on the number of expected background events, summarised in Table~\ref{tab:cut_flow}, and on Equation~(\ref{eq:sig_xsec}), the obtained limit is:
\begin{eqnarray*}
BR(\topqme) < 6.3 \cdot 10^{-5}   \;\;  \mathrm{@ 95 \% CL}.
\end{eqnarray*}
Alternative techniques for limit calculations, as implemented in RooStats, have been tried, leading to compatible results. 
An eventual variation of 100$\%$ in the number of expected background events would lead, in the worst case, to an expected limit of 
$BR(\topqme) < 7.4 \cdot 10^{-5}$ @ 95 \% CL. 

As explained in appendix~\ref{app:CMStZq}, we emulate in our framework 
the published CMS search for $BR(t\to Z q)$~\cite{CMS:tZc}. 
The reason for this exercise is twofold: on one hand it allows to validate our procedure on simulated samples, 
by comparing with the CMS background expectations. On the other hand, 
it  provides an estimate of the
constraint  set on  $BR(\topqme)$ from this previous  analysis, which is found to be $BR(\topqme)< 3.7\cdot 10^{-3}$, on the verge of probing 
LFV top decays mediated by a four-fermion operator.  
As proven in the present study, a dedicated analysis would set a limit 50 times stronger (of the order of $BR(\topqme) < 6.3 \cdot 10^{-5}$), 
showing that the existing LHC data from Run1 can still be used to obtain interesting constraints on lepton flavor violation.

\section{Discussion}
\label{sec:disc}

\subsection{Perspectives at 13 TeV and 14 TeV}
\label{ssec:13}

To estimate 
the reach of the described search  at a center-of-mass energy of  13 TeV,
we extrapolate
 the 8 TeV results, rather than performing a full simulation of signal and background processes at 13 TeV. 
The increase  of the  production cross sections for  SM processes, from 8 TeV to 13 TeV (see Table~\ref{tab:all_bkg_xsec}), is taken into account. 
The selection requirements and efficiencies are kept the same as for the 8 TeV analysis. For the signal, we have checked on simulated 
events that the efficiencies at 8 TeV and 13 TeV are consistent within 5$\%$.

The sensitivity is estimated by calculating the expected upper limits on $BR(\topqme)$, in the absence of signal, for two scenarios: the case of 
20 fb$^{-1}$ and 100 fb$^{-1}$ proton-proton data collected by the LHC at 13 TeV center-of-mass energy. 
We also extrapolate the sensitivity to the case of 3000 fb$^{−1}$ of integrated luminosity at 14 TeV. In this last case, we simply rescale signal and 
background rates from 13 TeV to 14 TeV, and use the square root of the number of expected background events as an estimate of their uncertainty. 
The obtained values are summarized in Table~\ref{tab:ul_13tev}. The upper limits presented here are derived using statistical uncertainties only, so don't 
take into account the possibility for such analyses to become systematically dominated in the future. In order to have an accurate evaluation of the 
systematics evolution, a deeper study from the LHC experiments would be needed. On the other hand, for an analysis on 13 TeV or 14 TeV data, the 
selection would have to be re-optimised, possibly leading to an increase in sensitivity.  

\begin{table} [h!]
\centering
\begin{tabular}{|c|c|c|c|c|}
\hline
    & 8 TeV (20 fb$^{-1}$) & 13~TeV (20 fb$^{-1}$) & 13~TeV (100 fb$^{-1}$) & 14~TeV (3000 fb$^{-1}$)\\
\hline
$BR(\topqme)$ & $< 6.3\cdot 10^{-5}$ & $< 2.9\cdot 10^{-5}$  & $< 1.2\cdot 10^{-5}$ & $\lappeq 2 \cdot 10^{-6}$ \\
\hline
\end{tabular}
\caption{Expected upper limits on $BR(\topqme)$, in the hypothesis of the absence of signal, for 8 TeV, 13 TeV (in two scenarios: the case 
of 20 fb$^{-1}$ and 100 fb$^{-1}$ collected luminosity) and 14 TeV for 3000 fb$^{-1}$ collected luminosity. \label{tab:ul_13tev}}
\end{table}

{ 
\subsection{Single Top}

In addition to mediating LFV top decays, 
the top operators listed in  Appendix~\ref{ssec:CIs}
could lead to single top production with an
 $e^\pm \mu^\mp$ pair, as illustrated in Figure~\ref{fig:afaire}. 
The objects in the final state  would be
the same as 
for the $t\bar{t} $ process we studied
: three leptons, missing energy,  and $\geq 2$ jets, of
which one is a $b$. We estimate\footnote{The  partonic diagrams for
$u+g \to e^- \mu^+ t$ are those
of
$b+g \to W+t$, with
$W $ replaced by par $e^- \mu^+$, 
and $b$ by $u$ or $c$. So we estimate
 $\sigma(pp \to  e^- \mu^+ t)$
by comparing to  $\sigma(pp \to  W t)
\sim 24$ pb $\sim 0.1\times \sigma (pp \to t\bar{t})$.
If we neglect spin correlations between
the $e^- \mu^+$ and $ ut$,
then  for
$V\pm A$ top operators, 
$|{\cal M} (ug \to  e^- \mu^+ t) |^2$
can be factorised as
$\propto \frac{1}{3}
|{\cal M} (ug \to  Z' t) |^2
\times
|{\cal M} (Z' \to  e^- \mu^+ ) |^2$
where the $Z'$ has coupling $1/\Lambda$
to fermions, and  the 1/3 is the spin average.
Then one has $|{\cal M} (Z' \to  e^- \mu^+ ) |^2 = 2 Q^2/\Lambda^2$
where $Q^2 = (p_e + p_\mu)^2$, and
$|{\cal M} (ug \to  Z' t) |^2 = 
2 |{\cal M} (bg \to  W t) |^2/(\Lambda^2 g^2)$.
The three body phase space integral can be
written as 
$\int d \Pi_{Wt} \times  d \Pi_{e \mu} \times dQ^2/(2\pi)$,
where  $d \Pi_{Wt}$ integrates the phase-space of
$W$ and $t$ in the centre-of-mass frame, 
 $ \int  d \Pi_{e \mu} = 1/(8\pi)$, and we
take  the upper limit of the $Q^2$ integral
as $m_W^2$, assuming thhis is a reasonable value
for leptons with $p_T> 20$ GeV.
Finally, at $m_t/(8$ TeV) $\lsim x\lsim  2m_t/8$  TeV, the
ratio of $u$ to $b$ pdfs  is of order 6-10, 
so we estimate that
$\sigma (pp \to e \mu^+ t) \sim 
\sigma(pp \to t\bar{t})\times BR(\topqme)$.
 } that  at 8 TeV, the cross section
for $pp \to e^\pm \mu^\mp  t\to  e^\pm \mu^\mp , \overline{\ell} \nu b$
is of similar order to 
the cross section
for $pp \to \bar{t} t  \to e^\pm \mu^\mp \bar{q} ,  \overline{\ell} \nu b$,
for operators involving a $u$ quark  and
slightly less for  a $c$ quark.

 \begin{figure}[t]
\unitlength.5mm
\SetScale{1.418}
\begin{boldmath}
\begin{center}
\epsfig{file=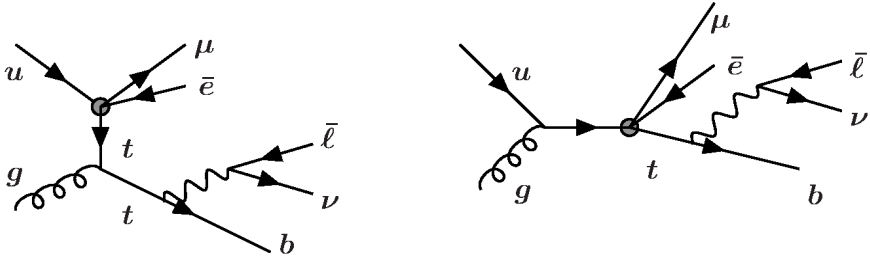, height=4cm,width=14cm}
\end{center}
\end{boldmath}
\caption{\label{fig:afaire}
Single top
production via the LFV contact  interaction, which 
produces a final state 
similar to the considered signal.
 }
\end{figure}

We neglect  this process for two reasons. First, 
the contact interaction approximation 
(that the $|$four-momentum$|^2$ in
the process $\ll \Lambda^2$)
is more difficult to justify
than in top decay, because the energy scale in the process
can be $\gg m_t$. Secondly, finding such events in the backgrounds
could be more challenging because  the $e^\pm \mu^\mp$
cannot be required to participate in the reconstruction of a top
(step 6 of Section~\ref{sec:sel}).

We envisage that it makes sense to neglect the LFV single
top process in  a first
search for LFV top decays. This is conservative,
because LFV  single top production could contribute
events that pass our selection.In the absence of a signal,
such a search could sufficiently constrain the contact interaction
scale $\Lambda$, to justify  including the
single top process in subsequent analyses.
}


\section{Summary}
\label{sec:sum}

The aim of this paper was to explore the LHC sensitivity 
to the decay $\topqme$, which is lepton and quark flavour-changing,
but baryon and lepton number conserving.
We parametrise this decay as
occuring via  a contact interaction,  and list a
complete set of   
$SU(3)\times U(1)$ invariant dimension-6  operators in
Appendix~\ref{ssec:CIs}. We parametrise the coefficient of these
interactions, which we refer to as ``top operators'', 
as $\epsilon 2\sqrt{2} G_F$ or equivalently $1/\Lambda^2$, with
$$
\epsilon \simeq \frac{m_t^2}{\Lambda^2}  ~~~.
$$ 
Model-building prejudice (see Section 
\ref{ssec:imp}), suggests that $\epsilon \lsim 1$.
The top branching ratio is then
$$
BR(\topqij) \simeq 1.3 \frac{|\epsilon|^2}{48\pi^2}
\times \left\{
\begin{array}{cc}
1 & V\pm A\\
\frac{1}{4} & S\pm P
\end{array}
\right. ~~~.
$$

These contact interactions are currently constrained from their
contribution in loops to rare  decays, and from single top
searches at HERA. 
   These bounds are discussed in the
Appendices, and summarised in section \ref{sec:2}.
For interactions involving $e$ and $\mu$, rare decay
bounds impose $\epsilon \lsim 0.01$  for some
operators, but others can have $\epsilon \sim 1$.

In Section~\ref{sec:experimental_section},
we  evaluate the sensitivity reach of a dedicated search 
for lepton flavor violation in top decays, at the 8 TeV LHC.
 The search targets $t\bar{t}$ events, where one top decays 
to an up-type quark ($u$ or  $c$) and a pair of leptons of 
opposite sign and opposite flavor, and the other one decays 
to a $b$ quark and a $W$, which subsequently decays to a 
charged lepton and a neutrino. This is illustrated
in Figure~\ref{fig:diagram_signal}.

The relevant signal and SM background processes are simulated 
for LHC Run1-like conditions: proton-proton collisions at 8 TeV 
center of mass energy,  for an integrated luminosity of about 20 fb$^{-1}$. 
The detector simulation is based on Delphes, with parameters tuned 
on the CMS detector reconstruction and performances, but 
 does not include pileup.  
The analysis setup is validated by emulating an existing CMS search 
for rare top decays to $Zq$ in $t\bar{t}$ events, showing reasonable results.

We find that a dedicated search by a single experiment using 
20 fb$^{-1}$  of  8 TeV data  could be sensitive to
$$BR(t \to e \bar{\mu} +  ~{\rm jet})  \sim 6.3\times 10^{-5}~~,$$
and extrapolate that  
 a sensitivity of  $ \sim 1.2 \times 10^{-5}$ ($ \sim 2 \times 10^{-6}$)
could be reached with 
 100 fb$^{-1}$ at 13 TeV (3000 fb$^{-1}$ at 14 TeV). 
From Equation (\ref{BRtop}), we see
that  the 100  fb$^{-1}$ data  could impose
$|\epsilon| \leq 0.06$ for $V \pm A$ operators,
and $|\epsilon| \leq 0.1$ for the $S \pm P$ and $LQ$ operators.
This analysis shows that the existing LHC data from Run1 can still be used to obtain interesting constraints on lepton flavor violation. 
Although this is understandably not the priority focus in the times of the Run2 startup, let's not to leave unchecked this possible path to New Physics.

\section*{Acknowledgements}
SD thanks P Gambino  and  Professor Y. Kuno for useful conversations,  and  
the labex OCEVU  and the theoretical physics
group at  Montpellier University-II for hospitality during
part  of this work.
The work of M.L.M.  is performed in the framework of the ERC grant 291377
``LHCtheory: Theoretical predictions and analyses of LHC physics:
advancing the precision frontier''.

\appendix

\section{Appendix: Operators}
\label{ssec:CIs}

Consider the $S$-matrix element mediating $\topqij$,
where $\ell_k \in \{ e,\mu, \tau\}$, 
and suppose it  is induced by local  operators 
with momentum-independent coefficients, such that they can be added
to the Standard Model Lagrangian.  These  operators
 should   respect 
the $SU(3)\times SU(2)\times  U(1)$ gauge symmetry
of the Standard Model. However, 
since the New Physics scale that we explore is
not much larger than the  electroweak scale $m_t$,
we should include dimension-8 operators
constructed from two Higgs fields and four fermions,
 or two gradients and four fermions.
Instead,
we choose   to work with $SU(3)\times U(1)$-invariant,
but {\it not} $SU(2)$-invariant,   operators of
dimension 6.
This is because  a dimension-8 $SU(2)$-invariant operator, 
can be a  dimension-6    $SU(3)\times U(1)$-invariant operator:
\beq
\frac{1}{\Lambda^4} 
(\overline{E}_e \gamma_\a E_\mu )
([\overline{Q}_2 H^c ] \gamma^\a [H^T i\tau_2 Q_3 ]) 
\to 
\frac{v^2}{\Lambda^4} 
(\overline{e} \gamma_\a P_R \mu )
(\overline{c} \gamma^\a P_L t) 
~~~,
\label{ex68}
\eeq
where $Q_i$ is a quark doublet of $i$th generation,
$E_j$ is a lepton singlet, $H^T= (H^+,H_0)$,
$H^c = i\tau_2 H^*$
and   SU(2) contraction  is in square brackets.

This choice of $SU(2)$-non-invariant
operators means that the
coefficients of $(\overline{e} \gamma_\a P_X \mu )
(\overline{s} \gamma^\a P_L b) $  and
 $(\overline{e} \gamma_\a P_X \mu ) \times$ $
(\overline{c} \gamma^\a P_L t)$
are taken as independent, and
in particular, bounds on the first
do not  apply directly to the second. 
However, dressing the
top operator with a $W$ loop will
generate the light quark vertex, 
which  gives unavoidable bounds 
that are estimated in  Appendix \ref{app:tloops}.
(Treating the $W$ as dynamical while
the New Physics is a contact interaction is unlikely to be
a good approximation, but  is the only way we can
estimate whether the top operators are in tension
with other observables.) In the case of $\meg$, 
 two or three
loops are required to transform the top operators
into the dipole operator, so we need  an SU(2)-invariant 
formulation of the top operators. These
dimension six and eight operators are
given in Appendix \ref{app:321}.

\subsection{$SU(3) \times U(1)$ invariant ``top operators''}

We are interested in contact interactions
involving  two  charged leptons of different flavour,
a top and a $c$ or $u$ quark.
Such colour-singlet,  electric charge-conserving,
 dimension-6  operators  
will be referred to as ``top operators''.  The $V\pm A$ operators are:
\bea
{\cal O}^{LL} &=&  (\overline{e}_i \gamma^\a P_L  e_j )
(\overline{u}_q \gamma_\a  P_L t ) 
\\
{\cal O}^{LR} &=&  (\overline{e}_i \gamma^\a P_L  e_j )
(\overline{u}_q \gamma_\a  P_R t ) 
\\
{\cal O}^{RL} &=&  (\overline{e}_i \gamma^\a P_R  e_j )
(\overline{u}_q \gamma_\a  P_L t ) 
\label{LR}
\\
{\cal O}^{RR} &=&  (\overline{e}_i \gamma^\a P_R  e_j )
(\overline{u}_q \gamma_\a  P_R t ) 
\eea
and  a redundant list of  scalar/tensor operators is:
\bea
{\cal O}^{S+ P,R}
 &=&   (\overline{e}_i P_R e_j ) 
(\overline{u}_qP_R t )  \\
{\cal O}^{S+ P,L}
 &=&   (\overline{e}_i P_L e_j ) 
(\overline{u}_qP_L t )  \\
{\cal O}^{S- P,R}
 &=&   (\overline{e}_i P_L e_j ) 
(\overline{u}_qP_R t )  \\
{\cal O}^{S- P,L}
 &=&   (\overline{e}_i P_R e_j ) 
(\overline{u}_qP_L t )  \\
{\cal O}^{T,R} &=&
(\overline{u}_q \sigma^{\mu\nu} P_R e_j )
(\overline{e}_i \sigma_{\mu\nu} P_R t )
\label{tensor}
\\
{\cal O}^{T,L} &=&
 (\overline{u}_q \sigma^{\mu\nu} P_L e_j )
(\overline{e}_i \sigma_{\mu\nu} P_L t )
\label{tensorL}
\\
{\cal O}^{LQ,R} &=& (\overline{u}_q P_R e_j )
(\overline{e}_iP_R t )
=-\frac{1}{2}{\cal O}^{S+ P,R}
 + \frac{1}{8}{\cal O}^{T,R} 
\label{LQTR}
\\
{\cal O}^{LQ,L} &=& (\overline{u}_q P_L e_j )
(\overline{e}_i P_L t )
=-\frac{1}{2}{\cal O}^{S+ P,L}
 + \frac{1}{8}{\cal O}^{T,L}
\label{LQTL}
\eea
where $e$, $u$ are Dirac spinors,
$i,j$ are unequal lepton flavour indices, 
$q \in \{u,c\}$,
and the last chiral superscript of the
operators
gives the chirality of the top.

The scalar operators ${\cal O}^{LQ,X}$ 
can be exchanged for 
 the tensor operators  ${\cal O}^{T,X}$,
as shown in Equations 
(\ref{LQTR}) and (\ref{LQTL}).
We will use  the $LQ$ and $T$ operators interchangeably, becuase 
the ${\cal O}^{LQ,X}$ 
are more convenient in  top decay,  and 
the tensors at low
energy.
The tensors are used in the basis of~\cite{polonais}.
Notice that $S+P$ operators are defined to have the
same chiral projector twice,  whereas for $S-P$
they  are  different.
Since  the last  quark flavour label is fixed to $t$,
 $S + P$, $S-P$ and $LQ$ operators appear
twice,   
for  both chiralities   $t_L$ and $t_R$
(were we using arbitrary quark flavour indices $q,v$,
then $[{\cal O}^{S+ P,R}_{ijqv}]^\dagger = {\cal O}^{S+ P,L}_{jivq}$).
Finally, the scalar  operator contracting
quark  to lepton spinors  is given the name $LQ$;
there are no such operators of the form
$(\overline{u}_q P_R e_j )
(\overline{e}_j P_L u_t )$, because this
is Fierz-equivalent to  ${\cal O}^{RL}$.

A similar list can be constructed for down-type
quarks. These will be relevant, because
the top-operators can generate the down-operators
at one loop.  To reduce index confusion, down-type
quarks will have greek flavour indices $\alpha,\beta
\in \{d,s,b\}.$ There are also  charge-current (CC) scalar operators
\bea
{\cal O}^{CC} =
(\overline{\ell}_i  P_L \nu_j)(\overline{u}_q  P_L d_\beta)
~~~~,~~~~~~
{\cal O}^{CC,LQ} = 
(\overline{u}_q  P_L  \nu_j)(\overline{\ell}_i  P_L d_\beta)
\label{CC}
\eea
which will  be relevant in setting bounds.

These  operators
are included in the Lagrangian
with a    dimensionful coefficient
\beq
-\frac{ 1}{\Lambda^2}  \equiv -2 \sqrt{2} G_F \epsilon ~~.
\eeq 
The $\epsilon$s have subscripts that identify the
operator, and lepton-quark flavour superscripts: $2 \sqrt{2} G_F 
\epsilon_{LL}^{ijqt}$
would be the coefficient of $ (\overline{e}_i \gamma^\a P_L  e_j )
(\overline{u}_q \gamma_\a  P_L t ) $,
and $\epsilon_{LQ,X}^{ijqt}$
would be the coefficient of $ (\overline{u}_q P_X  e_j )
(\overline{e}_i  P_X t ) $ (so in particular,
the index order is always $\bar{\ell} \ell \bar{q} q$,
even for ${\cal O}^{LQ}$).

\subsection{$SU(3) \times SU(2) \times U(1)$ operators}
\label{app:321}

Our  $V\pm A$ operators can be  identified
as  components of 
SU(2) invariant  dimension 6 operators:
\bea
{\cal O}^{LL} &=&  (\overline{e} \gamma^\a P_L  \mu )
(\overline{u}_q \gamma_\a  P_L t ) 
\subset  {\cal O}_{L Q}^{(1)} = (\overline{L}_e \gamma^\a L_\mu ) 
(\overline{Q}_q \gamma_\a Q_t ) \label{meg12} \\
&&~~~~~~~~~~~~~~~~~~~~~~~~~~~~~~~~~~~~ = [(\overline{e} \gamma^\a P_L  \mu )
+ (\overline{\nu}_e \gamma^\a P_L  \nu_\mu )]
[(\overline{u}_q \gamma_\a \tau^a P_L t ) 
+ (\overline{d}_q \gamma_\a \tau^a P_L b )]
\nonumber\\
{\cal O}^{LR} &=&  (\overline{e} \gamma^\a P_L  \mu )
(\overline{u}_q \gamma_\a  P_R t ) 
\subset {\cal O}_{LU} = (\overline{L}_e \gamma^\a L_\mu ) 
(\overline{U}_q \gamma_\a U_t ) 
\label{meg13}\\
{\cal O}^{RL} &=&  (\overline{e} \gamma^\a P_R  \mu )
(\overline{u}_q \gamma_\a  P_L t ) 
\subset
{\cal O}_{EQ} = (\overline{E}_e \gamma^\a E_\mu ) 
(\overline{Q}_q \gamma_\a Q_t ) \label{meg14}\\
{\cal O}^{RR} &=&  (\overline{e} \gamma^\a P_R  \mu )
(\overline{u}_q \gamma_\a  P_R t ) 
={\cal O}_{EU} = (\overline{E}_e \gamma^\a E_\mu ) 
(\overline{U}_q \gamma_\a U_t )
\label{meg15}
\eea
where $L,Q$ are doublets and $E,U$ are singlets
$t$ is a third generation quark index, 
$q$ a first or second generation quark index,
and we explicitly choose lepton indices $i =e$
and $j = \mu$.
In the SU(2) invariant list, 
there is  an additional operator
at dimension six, which  generates ${\cal O}_{L L}$ 
in combination with  charged current interactions:
\bea
{\cal O}_{L Q}^{(3)} &=& (\overline{L}_i \gamma^\a \tau^aL_j ) 
(\overline{Q}_q \gamma_\a \tau^a Q_t ) \nonumber \\
&=&  2 (\overline{\nu}_i \gamma^\a P_L  e_j )
(\overline{d}_q \gamma_\a  P_L t ) 
+ 2 (\overline{e}_i \gamma^\a P_L  \nu_j )
(\overline{u}_q \gamma_\a  P_L b ) \nonumber\\ 
&&~~~~~+ [(\overline{\nu}_i \gamma^\a P_L  \nu_j ) -
(\overline{e}_i \gamma^\a P_L  e_j )]
[(\overline{u}_q \gamma_\a  P_L t ) 
- (\overline{d}_q \gamma_\a  P_L b )]~~~.
 \label{LQ3}
\eea

Alternatively, at dimension eight, SU(2) invariant
operators can be constructed to contain only
an up-type quark current,
by contracting quark and Higgs  doublets.
An example is  given in Equation (\ref{ex68}).

Then there are two   $S + P$ operators:
\bea
{\cal O}^{S+ P,R}
 &=&   (\overline{e} P_R \mu ) 
(\overline{u}_qP_R t )   
\subset
{\cal O}_{L_1 E_2 Q_2 U_3} \nonumber \\
{\cal O}^{S+ P,L}
 &=&   (\overline{e} P_L \mu ) 
(\overline{u}_qP_L t )    
\subset {\cal O}_{L_2E_1 Q_3 U_2}^\dagger \nonumber 
\eea
 which
are  components of the same SU(2) invariant
operator:
\bea
{\cal O}_{LE Q U}&=& (\overline{L}_i^A  E_j )\epsilon_{AB} 
(\overline{Q}^B_q U_t )
~~~~~~~~~~~~~~~~~~~~~~~~
\label{scalar}\\
 &=& -(\overline{\nu}_i  e_j ) 
(\overline{d}_q u_t ) + (\overline{e}_i P_R e_j ) 
(\overline{u}_qP_R u_t ) \nonumber 
\eea
and there are the LQ or tensor
operators,
which are components of the   dimension six,  SU(2)-invariant operator
\bea
{\cal O}_{LUQE} &=& (\overline{L}_i^A   U_t )
\epsilon_{AB}  (\overline{Q}^B_q E_j )
=-\frac{1}{2}
{\cal O}_{LE Q U} + \frac{1}{8} (\overline{L}_i^A \sigma^{\mu\nu} E_j )
\epsilon_{AB} 
(\overline{Q}^B_q \sigma_{\mu\nu} U_t ) ~~~,
\label{tensorSU2}
\eea
and finally there are
  two $S-P$ operators,
which can be identified 
with  an SU(2)-invariant operator of  dimension eight:
\bea
v^2 {\cal O}^{S- P,R}
 &=& v^2  (\overline{e} P_L \mu ) 
(\overline{u}_qP_R t )  \\
& =  & (\overline{E}_1  [H^\dagger   L_2] ) 
([\overline{Q}_q  H^c ]  U_3 )  \nonumber \\
v^2 {\cal O}^{S- P,L}
 &=  &   ([\overline{L}_e H ] E_\mu ) 
(\overline{U}_q [H^T i\tau_2 Q_t] )  
\eea
where, for instance, $ [H^\dagger   L_2] = v \mu_L$.
Notice that our  
$S+P$ and tensor  operators can also be
obtained at dimension 8 without 
 associated charged current interactions:
\beq
(\overline{L}_i H  E_j ) 
(\overline{Q}_q H^c U_t )
=  v^2 (\overline{e}_i P_R e_j ) 
(\overline{u}_qP_R u_t ) 
\eeq

\section{Appendix: top operators  in $\meg$}
\label{app:meg}

In this Appendix, we wish to close the quark lines
of a top operator, 
and attach a photon  (and a Higgs vev) to the resulting diagram, such that it
contributes to $\meg$. These two and three loop
estimates rely on using  SU(2)-invariant operators (of
dimension six or eight), because using the  $d_L$ and
$\nu_L$  components of doublets will allow to avoid
GIM suppression. Then we  estimate the
contribution  to $\meg$ 
by power counting selected diagrams.

\subsection{Parametrising $\meg$}

The decay $\meg$ is mediated by the dipole operator,
which can be  added to the Standard Model Lagrangian  as
\cite{LFV}
\bea
{\cal L}_{meg} &= &-\frac{4G_F}{\sqrt{2}} m_\mu \left(
A_R \overline{\mu_R} \sigma^{\a\b} e_L F_{\a\b} +
A_L \overline{\mu_L} \sigma^{\a\b} e_R F_{\a\b}\right)\nonumber \\
&=&
 -\frac{4G_F}{\sqrt{2}} y_\mu \left(
A_R \overline{\mu_R} \sigma^{\a\b} [H^\dagger L_e] F_{\a\b} +
A_L [\overline{L_\mu} H] \sigma^{\a\b} e_R F_{\a\b}\right)
\label{Lmeg}
\eea
where, in the second formulation, $y_\mu$ is
the muon Yukawa coupling such that 
$y_\mu v= m_\mu$, and 
the fermion part has  been  written in SU(2)-invariant
form, to emphasize that the dipole operator
is of dimension six, and has a Higgs leg. 
This operator gives a  branching ratio:
\beq
BR(\meg)= 384 \pi^2 (|A_R|^2 + |A_L|^2) <5.7\times 10^{-13}
\label{BRmeg}
\eeq
where we quote the upper bound from \cite{MEG}.
If $|A_R| = |A_L|$, this implies $|A_X| <8.6\times 10^{-9}$.
However, in the case of our diagrams, there
will be a sum over three colours at the amplitude-squared
level, so its convenient to divide by
$\sqrt{3}$, and impose
\beq
|A_X| <5.0 \times 10^{-9}.
\label{AXmeg}
\eeq
This is a small number.  
We can estimate the  contribution  of  top operators
to be
suppressed by various factors:
\ben
\item  two loops, and in some cases three. 
Two loops are neccessary to close  the quark legs,  
because they  are of different flavour, 
so must interact with a $W$.
\item CKM, because the quark
legs are of different generation. This is $\sim .04$ for $c\leftrightarrow t$
flavour change, $\sim .008$ for $u \leftrightarrow t$
flavour change.
\item GIM: if the W is exchanged between two
up-type
quark lines (see Figure \ref{fig:megtopop}), then
the diagram is further suppressed by a GIM
factor  $\sim m_b^2/m_W^2 \simeq 1/400$ (for $m_b= 4$ GeV).
This can be avoided by exchanging the $W$ between
the quark and lepton lines.
\item mass insertions: may be required on the quark line
to transform singlet into doublet quarks. However, the
dipole operator of Equation (\ref{Lmeg}) is defined with a Higgs leg
attached to the muon line;   so this can be an
amplification factor if the Higgs 
can be attached to a heavier fermion.
\een
So for our $SU(3)\times U(1)$
invariant  top operators with coefficient $\epsilon \frac{4G_F}{\sqrt{2}}$,
  one  could anticipate 
\beq
A_X \sim  \frac{ e g^2/2}{(16\pi^2)^2} \epsilon 
\times CKM \times GIM  \times m\!-\!insertations \simeq 
\left\{
\begin{array}{rlc}
  10^{-7} &
\epsilon 
 \times {GIM} \times m\!-\!insertations 
&~,~ q = c\\
  2\times 10^{-8} &
\epsilon 
 \times {GIM} \times m\!-\!insertations 
&~,~ q = u
\end{array}
\right.
\eeq
where, in  some diagrams, the GIM  factor is absent,
and/or the mass insertion factor can be $\gg 1$.
Comparing to Equation (\ref{AXmeg}) suggests that
$\meg$ can provide relevant constraints on the
$\epsilon$s, unless  there is more  suppression 
than CKM and two loops.

\subsection{Estimated contributions to $\meg$}

It is convenient to pretend that
the New Physics scale is $\gg v$,
which allows to study
 whether  (the SU(2) invariant generalisations
of) the top operators  mix to the dipole  in 
electroweak running
down to $v$. The aim is to add
electroweak boson loops  to the top
operator such that the quark lines can
be closed, and  the dipole operator
of Equations (\ref{Lmeg}) is obtained. Closing
the quark lines requires  changing  the quark flavours. 
The  lepton spinor contraction of
the dipole is tensorial, so 
for all non-tensor top operators, 
it must be changed to this form by 
involving a lepton line in a loop. 

\ben
\item Consider first  the SU(2) invariant
versions of  the $V\pm A$ operators of 
Equations (\ref{meg12}-\ref{meg15}). These operators
have no $(\bar{e} \g \nu)(\bar{u} \g d)$ components,
so two $W$ vertices are required on the quark loop
(see  the  first diagram of  figure \ref{fig:megtopop}).
This gives 2-loop  and  CKM suppression.
For singlet quark currents, also
mass insertions are required on the $t$ and $q$ lines,
and the diagram will be suppressed by
a GIM factor $\sim m_b^2/m_W^2$ (these
mass factors correspond to the crosses
on the quark loop of the first diagram of Figure
\ref{fig:megtopop}).
  To modify the spinor contraction
between the leptons, 
a $\g,Z$  could be exchanged between the  quark loop
and an external lepton line,
which gives an additional suppression
 $\sim e^2/(16\pi^2)$. So for singlet quark
currents (the operators of Equations (\ref{meg13})
and (\ref{meg15})): 
\beq
A_X \sim  \frac{ e^3 g^2/2}{(16\pi^2)^3} \epsilon 
\times V_{qb}  \frac{m_b^2}{m_W^2} \frac{m_q}{m_t}   \simeq   10^{-15} 
\epsilon 
\times \frac{V_{qb}}{.04} 
\eeq
and for doublet quark currents, where the  GIM and $m_q$  suppression
can be avoided by emitting $d$-type quarks from
the operator:
\beq
A_X \sim  \frac{ e^3 g^2/2}{(16\pi^2)^3} \epsilon 
\times V_{tq}    \simeq    10^{-10} 
\epsilon 
\times \frac{ V_{tq} }{.04} 
\eeq
These loops gives no bound  on  $\epsilon_{LL}$,
$\epsilon_{LR}$, $\epsilon_{RL}$ or $\epsilon_{RR}$.

\begin{figure}[h]
\begin{center}
\epsfig{file=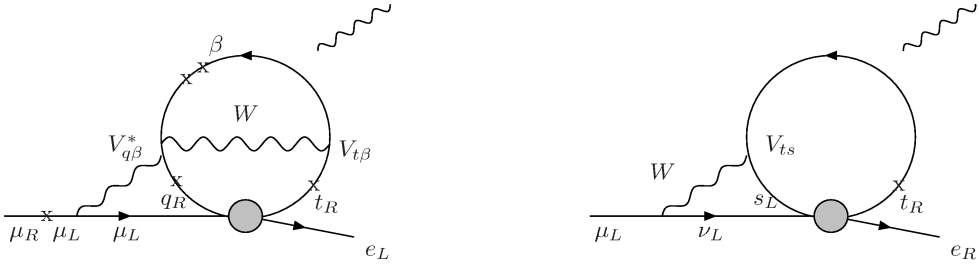, height=4cm,width=14cm}
\end{center}
\caption{Left: example  three-loop diagram by which a
the  neutral current vector operator of
Equation (\ref{meg13})  could
induce $\meg$.  ``x'' is a mass insertion, the grey
blob is the  operator, and the photon
could be attached  to any
quark line.  On the right is an
 example  two-loop diagram by which  
a tensor/scalar  operator  could
induce $\meg$ without GIM suppression and with
$m_t/m_\mu$ enhancement.
 \label{fig:megtopop}
}
\end{figure}

\item However, if  the ${\cal O}_{LL}$ operator
is generated as a component of the triplet SU(2)-invariant
operator of Equation  (\ref{LQ3}), then  there is a charged
current contact interaction, which allows $W$ exchange
between the lepton and quark lines and
avoids any GIM suppression. The  diagram
is similar to 
Figure   \ref{fig:megtopop} on the right,
but with $t_L$ and $e_L$ instead of 
 $t_R$ and $e_R$, and a mass insertion on
the external $\mu$  rather than  the $t$.
An estimate gives
\beq
A_X \sim  \frac{ e g^2/2}{(16\pi^2)^2} \epsilon 
\times CKM     \simeq   10^{-7} 
\epsilon 
\times \frac{CKM}{.04} 
\eeq
which imposes  $\epsilon_{LL} \lsim 0.03$ for $q=c$,
or $\epsilon_{LL} \lsim 0.5$ for $q=u$.
 Note that the B decay constraints, which are estimated 
below more rigorously than these ones, are tighter.

\item  Consider now the $S+P$ or $T$  operators which
arise at dimension  six, which can induce
the second diagram 
of Figure \ref{fig:megtopop}.  The grey
blob represents the charge-changing component of the  operator, 
 involving $t_R$ and either $d_L$ or
$s_L$.  The contribution
to $\meg$  is two-loop and
CKM suppressed, but enhanced by
$m_t/m_\mu$, because the chirality flip
for the leptons is in the contact interaction,
so the Higgs leg of the dipole operator can attach
to the mass insertion required to
flip the chirality of the top. This gives
\beq
A_X \sim  \frac{ e g^2/2}{(16\pi^2)^2} \epsilon 
\times 
\left\{
\begin{array}{c}
V_{ts} \\
V_{td}
\end{array}
\right\}
 \times\frac{m_t}{m_\mu}     \simeq  
\left\{
\begin{array}{rl} 2 \times 10^{-4} 
&\epsilon^{ct} \\
6 \times 10^{-5} 
&\epsilon^{ut} 
\end{array}
\right.
\eeq
which  implies $\epsilon_{S+P,R}, \epsilon_{T,R}  \lsim 10^{-4}\to  10^{-5}$! 

In the case   where  the third generation quark
is  a doublet,
then there would be a $b_L$ and a charm or up mass
insertion, so that
\beq
A_X \sim  \frac{ e g^2/2}{(16\pi^2)^2} \epsilon 
\times  V_{qb}   \frac{m_q}{m_\mu} \sim 
\left\{
\begin{array}{rl}
 10^{-6} &
\epsilon^{ct} \\
 6 \times 10^{-10}& 
\epsilon^{ut}
\end{array}
\right.
\eeq
which would suggest  $\epsilon_{S+P,R}^{ct}, \epsilon_{T,R}^{ct}  \lsim 10^{-3}$.

\item Finally, consider   an $S\pm P$  or $T$ operator of  
dimension eight, so with two Higgs legs.  If the Higgs are both
neutral vevs, then the contact interaction
involves up-type quarks, and gives diagrams
whose contribution
to $\meg$ is suppressed by three loops, CKM and GIM.
Instead, focus  on diagrams where the Higgs
legs are charged, so the four-fermion interaction
is charge-changing for the leptons and quarks,
and the Higgs loop has to close (two external Higgs 
legs can be attached  to the top line, to
keep the operator at dimension eight, 
which gives negligible suppression because $y_t \sim 1$). 
Due to the Higgs loop, the estimate for 
$A_X$   will be suppressed by $\sim 10^{-2}$
 with respect
to the previous  discussion of $S+P$ and $T$
at dimension  six. Since we do not
know whether our ``top operators''
correspond to SU(2) invariant operators
of  dimension six or eight, we conservatively
retain the dimension eight bounds, 
which imply that for  $S\pm P$  and $T$ operators
involving  a singlet $t_R$:
\beq
\epsilon_{T,R}^{ct}~,~
\epsilon_{S\pm P,R}^{ct} \lsim 10^{-3} ~~~,~~~
\epsilon_{T,R}^{ut}~,~
\epsilon_{S\pm P,R}^{ut} \lsim 10^{-2} ~~.
\label{megbds}
\eeq

\een

Similar estimates  can be made for the  contributions
of top operators to 
$\tau\to \ell \gamma$. The
current experimental upper bounds \cite{tlg} are
$\Gamma(\tau\to e \gamma) < 1.8\times 10^{-7}
\Gamma(\tau\to e\nu\bar{\nu})$ and
$\Gamma(\tau\to \mu \gamma) < 2.5 \times 10^{-7}
\Gamma(\tau\to \mu \nu\bar{\nu})$, which imply,
with the assumptions leading to Equation
(\ref{AXmeg}), that $A_X(\tau \to \ell \g)  \lsim 3\times 10^{-6}$.
For dimension eight, $S\pm P$ or tensor operators,
this would give $\epsilon^{\ell\tau ut} \lsim 100$
and $\epsilon^{\ell\tau ct} \lsim 10$.

\section{Appendix: top operators  in meson decays}
\label{app:tloops}

Dressing a top operator with a 
$W$ loop, where the $W$ attaches to the top leg, 
generates a
contact interaction among light fermions,
where the $t$ is replaced by $b$,$s$ or $d$.
So the first step in obtaining bounds on 
the top operators, is to collect up-to-date 
bounds on the relevant two-quark-two lepton
operators invoving $d$-type quarks.  These
are listed in Tables~\ref{tab:eedd}, \ref{tab:SP}
and \ref{tab:SPnu}, updated 
from~\cite{Carpentier}. In Section~\ref{ssec:loops} , we
estimate the loop  factors,  which,
when 
multiplied by the top-operator coefficients,
give the coefficients of light-fermion operators
that are constrained by Tables~\ref{tab:eedd}, \ref{tab:SP}
and \ref{tab:SPnu}. Finally,
in Section~\ref{ssec:bounds}, we combine
the loop factors and Tables to obtain the
bounds on top operator coefficients given in
Tables~\ref{tab:topVA} and \ref{tab:topSP},
and briefly discuss possible cancellations
among the operators.

\subsection{ Updated rare decay bounds }
\label{ssec:raredec}

\begin{table}[htp!]
\begin{center}
\begin{tabular}{|c|c|c|c|c|}
\hline
&&&&\\
$ij\a \b$
 & 
&
 Constraint on $\epsilon^{ij \a \b}$ & Observable &Experimental value\\
&&&& \\
\hline
\hline
$e\mu ds$   
&$(\epsilon_{LL}  - \epsilon_{LR}) ~,~ (\epsilon_{RR} -\epsilon_{RL})$
& $3.0\times10^{-7}$   
& $BR(K^{0}_{L}\rightarrow \bar{e}\mu)$ 
& $<4.7\times10^{-12}$\\
\hline
$e\mu db$ 
&$(\epsilon_{LL}  - \epsilon_{LR}) ~,~ (\epsilon_{RR} -\epsilon_{RL})$
& $3.0\times10^{-4}$     
&$BR(B^{0}\rightarrow \bar{e}\mu)$ 
& $<2.8\times10^{-9}$\\
&$\epsilon_{LL}  +\epsilon_{LR} ~,~ \epsilon_{RR} +\epsilon_{RL}$
& $1.3\times10^{-4}$     
&$\frac{BR(B^{+}\rightarrow\pi^+ \bar{e}\mu)}
{BR(B^{0}\rightarrow\pi^- \bar{e}\nu)}$ 
& $<1.3\times10^{-3}$\\
\hline
$e\mu sb$ 
&$\epsilon_{LL}  +\epsilon_{LR} ~,~ \epsilon_{RR} +\epsilon_{RL}$
& $1.0 \times10^{-4}$       
& $\frac{BR(B^{+}\rightarrow K^{+}\bar{e}\mu)}
{{BR(B^{+}\rightarrow D^{0}\bar{e}\nu_{e})}}$ 
& $<\frac{9.1\times10^{-8}}{2.2\times10^{-2}}$\\
\hline
$e\tau ds$ 
&$(\epsilon_{LL}  - \epsilon_{LR}) ~,~ (\epsilon_{RR} -\epsilon_{RL})$
& $4.1\times10^{-4} $    
& $\frac{BR( \tau \rightarrow {e} K)}
{BR( \tau \rightarrow \bar{\nu} K)}$ & $ <\frac{2.6  \times10^{-8}}
{7.0 \times10^{-3}}$\\
\hline
$e\tau db$ 
&$(\epsilon_{LL}  - \epsilon_{LR}) ~,~ (\epsilon_{RR} -\epsilon_{RL})$
& $2.1\times10^{-3}$    
& $BR(B^{0}\rightarrow \bar{e}\tau)$ 
& $<2.8\times10^{-5}$\\
\hline
$e\tau sb$
&$\epsilon_{LL}  +\epsilon_{LR} ~,~ \epsilon_{RR} +\epsilon_{RL}$
& $2.0  \times10^{-3}$       
& $\frac{BR(B^+ \to K^+ \bar{\tau} e)}
{{BR(B^{+}\rightarrow D^{0}\bar{\tau}\nu)}}$ 
& $<\frac{3.0\times10^{-5}}{7.7\times10^{-3}}$\\
\hline
$\mu \tau ds$ 
&$(\epsilon_{LL}  - \epsilon_{LR}) ~,~ (\epsilon_{RR} -\epsilon_{RL})$
& $4.3\times10^{-4} $    
& $\frac{BR( \tau \rightarrow {\mu} K)}
{BR( \tau \rightarrow \bar{\nu} K)}$ & $
 < \frac{2.3  \times10^{-8}}
{7.0 \times10^{-3}}
 $\\
\hline
$\mu\tau db$ 
&$(\epsilon_{LL}  - \epsilon_{LR}) ~,~ (\epsilon_{RR} -\epsilon_{RL})$
& $1.6\times10^{-3}$  
& $BR(B^{0}\rightarrow \bar{\mu}\tau)$ 
& $<2.2\times10^{-5}$\\
\hline
$\mu \tau sb$ 
&$\epsilon_{LL}  +\epsilon_{LR} ~,~ \epsilon_{RR} +\epsilon_{RL}$
& $3.1  \times10^{-3 }$       
& $\frac{BR(B^+ \to K^+ \bar{\tau} \mu)}
{{BR(B^{+}\rightarrow D^{0}\bar{\tau}\nu)}}$ 
& $<\frac{4.8\times10^{-5}}{7.7\times10^{-3}}$\\
\hline
\end{tabular}
\caption{
Constraints on the dimensionless
coefficient  $\epsilon^{ij\a \b}$, of  four-fermion
interactions $2 \sqrt{2}  G_F \left(\bar{e}_{i} \gamma^{\mu} P_{L,R}
e_{j}\right)\left(\bar{d}_{\a}\gamma_{\mu} P_{L,R} d_{\b}\right) $. 
 $P_{L,R}$ can be $P_{L}$ or $P_{R}$.
The 
generation indices $ij\a \b$  are given in the first column, 
and the  best bound   in 
column three. It  arises from  the observable of column 4,
and  the experimental value we used is given  in column 5.
In column 2, is given the linear combination of
$\epsilon$s to which the bound applies.
All bounds apply under permutation of the lepton and/or
quark indices.
}
\label{tab:eedd}
\end{center}
\end{table}

The bounds
from the  decay  rate of a pseudoscalar
meson $M$,
 are obtained using the 
usual  formula:
\bea
 \Gamma \left(M \rightarrow l^{i}\bar{l}^{j}\right)
& =& \frac{kG_F^2}{\pi m^{2}_{M}} 
\Bigg\{ \widetilde{P}^{2} 
\left[
 (|\epsilon_{S \pm P,R} - \epsilon_{S\pm P,L}| ^{2})
\left(m^{2}_{M}-m^{2}_{i}-m^{2}_{j}\right) 
+ 4 \epsilon_{L} \epsilon_{R} m_1 m_2 \right] 
+  \nonumber\\   &&( |\epsilon_{XR} - \epsilon_{XL}|^{2})
\widetilde{A}^{2}\left[(m^{2}_{M}-m^{2}_{i}-m^{2}_{j})(m^{2}_{i}+m^{2}_{j}) +4m^{2}_{i}m^{2}_{j}\right]
\Bigg\} ~,
\label{8}
\end{eqnarray}
where  scalar-vector interference was neglected.
In current algebra, the  quark currents  are:
\beq
\widetilde{A}P^\mu= \frac{1}{2} 
\langle 0| \overline{q} \gamma^\mu \gamma^5 q |M \rangle 
=  \frac{f_{M}P^\mu}{2}  ~~~~~~~
\widetilde{P}=  \frac{1}{2} 
\langle 0| \overline{q}_o  \gamma^5 q_n |M \rangle 
= \frac{f_{M}m_{M}}{2}\frac{m_{M}}{m_{o}+m_{n}}~~,
\label{currentalgebra}
\eeq 
where  $o,n$ are flavour indices,  and 
$k$ is the magnitude of the final state three-momenta in the centre-of-mass
frame: 
	\begin{equation}
k^{2}=\frac{1}{4m^{2}_{M}}\left[\left(m^{2}_{M}-\left(m_{i}+m_{j}\right)^{2}\right)\left(m^{2}_{M}-\left(m_{i}-m_{j}\right)^{2}\right)\right]~~~.
\label{9}
\end{equation}

The  decay $B^+\to K^+ e^-\ell^+$
can be used to constrain  $\epsilon^{e \ell  sb}$,
with a suitable approximation for the
hadronic matrix element.  For this, we follow the
 Babar exclusive determination of $V_{cb}$~\cite{VcbBabar}
from  
$B^+\to D^0 \ell^+\nu$,
where  the differential decay
rate is parametrised with a a kinematic term depending
on $q^2= (p_B - p_D)^2$ and a form factor, that is function of
the inner product $\o$  of $B$ and $D$ 4-velocities:
$$
\frac{d \Gamma}{d \o} = 
\frac{G_F^2|V_{cb}|^2m_B^3}{8 \times 48\pi^3}(m_B+m_D)^2 (1 + {\cal O}(m_D^2/m_B^2) +...){\cal G}^2(\o)
$$
where $\dots$ includes the $q^2$ terms.
To extract a bound on $\epsilon$ from
$B^+\to K^+ e^-\tau^+$, we  suppose the
form factors are similar, and impose
\beq
\frac{\Gamma(B^+\to K^+ e^-\tau^+)}
{\Gamma (B^+\to D^0 \ell^+\nu)} = \frac{\epsilon^2}{|V_{cb}|^2}
 \frac{(m_B+m_K)^2}{(m_B+m_D)^2} ~~~,
\label{BDK}
\eeq
which is slightly weaker than the bound in~\cite{Carpentier},
for the case $\ell = \mu$.

For $S\pm P$ operators, 
we  included $\tau \to K_S \ell$
bounds, that were not in~\cite{Carpentier},
but used in~\cite{KOT}.

\begin{table}[!tpb]
\begin{center}
\begin{tabular}{|c|c|c|c|}
\hline
&&&\\
$ij\alpha \beta$ & Constraint on $\epsilon^{ij\a \b}$ & Observable & Experimental value\\
&&& \\
\hline
\hline
$\mu eds$    & $9.0\times10^{-9}$   & $BR(\overline{K^{0}_{L}}\rightarrow \bar{\mu}e)$ & $<4.7\times10^{-12}$\\
\hline
$e\mu bd$  &  $8.1\times10^{-6}$& $BR(B^{0}\rightarrow \bar{e}\mu)$ & $<9.2\times10^{-8}$\\
\hline
$e\mu bs$  & $2.9\times10^{-5}$ & $BR(B^{0}_{s}\rightarrow \bar{e}\mu)$ & $<1.1\times10^{-6}$\\
\hline
$e\tau ds$   &
$2.9\times10^{-4}$ & $BR(\tau \rightarrow K_S^0 e)$ 
& $<2.6\times10^{-8}$\\
\hline
$e\tau db$  & $5.6\times10^{-4}$ & $BR(\overline{B^{0}}\rightarrow \bar{e}\tau)$ & $<2.8\times10^{-5}$\\
\hline
$e\tau sb$   &&& \\
\hline
$\mu\tau bd$  & $5.0\times10^{-4}$ & $BR(B^{0}\rightarrow \bar{\mu}\tau)$ & $<2.2\times10^{-5}$\\
\hline
$\mu\tau ds$ &
$2.6\times10^{-4}$ & $BR(\tau \rightarrow K_S^0 e)$ 
& $<2.3\times10^{-8}$\\
\hline
$\mu\tau sb$  & &&\\
\hline
\end{tabular}
\caption{ Constraints  on
$ \epsilon_{S \pm P,X} $,
for the $ij \a \b$ index combination given in the first column,
where the $\epsilon$s are coefficients of the operators $ {\cal O}_{S \pm P,X}=
\left(\bar{e}_{i}P_Y e_{j}\right)\left(\bar{d}_{\a}P_Xd_{\b}\right)$,
in a basis containing also tensor operators.
The second column is the constraint, which
 arises from the observable  given  in column 3.
The  experimental value  used is the last column.
These bounds  are also valid under lepton  
and/or quark index permutation.}
\label{tab:SP}
\end{center}
\end{table}

\begin{table}[!tpb]
\begin{center}
\begin{tabular}{|c|c|c|c|}
\hline
&&&\\
$\left(\bar{\nu}_{i}{e}_{j}\right)\left(\bar{d}_{\beta}u_{q}\right)$ & Constraint on $\epsilon^{ij\beta q}$ & Observable & Experimental value\\
&&& \\

\hline
\hline
$\nu_{i}edu$  & $1.6 \times10^{-5}$   & $R_{\pi}$& $(1.230\pm 0.004)\times10^{-4}$\\
\hline
$\nu_{i}edc$  &   $1.4 \times10^{-3}$   & $\Gamma(D^+ \to \bar{e} \nu_i)$& 
$< 8.8 \times10^{-6}$\\  
\hline
$\nu_{i}e su$& $1.5\times10^{-5}$ & $BR(K^+ \to \bar{e} \nu_i)$ 
 & $(1.55\pm 0.07)\times10^{-5}$\\
\hline
$\nu_{i}esc$  & $ 3.9 \times 10^{-3}$     & $BR(D_s^+ \to \bar{e} \nu_i)$   
& $ < 8.3 \times 10^{-5} $ \\
\hline
$\nu_{i}ebu$   & $7.8\times10^{-5}$ & $BR(B^{+}\rightarrow \bar{e}\nu)$ & $<9.8\times10^{-7}$\\
\hline
$\nu_{i}ebc$  &     &  & \\
\hline
$\nu_{i}\mu du$   & $3.2 \times10^{-3}$ & $R_{\pi}$& $(1.230\pm 0.004)\times10^{-4}$\\
\hline
$\nu_{i}\mu dc$  &  $3.7 \times10^{-3}$   
& $BR(D^{+}\rightarrow \bar{\mu}\nu)$ & $(3.82\pm 0.33)\times10^{-4}$\\ 
\hline
$\nu_{i}\mu su$   & $3.0\times10^{-3}$ & $R_{K}$& $(2.44\pm 0.11)\times10^{-5}$\\
\hline
$\nu_{i} \mu sc$  & 
 $1.5\times10^{-2}$   
& $
{BR(D^{+}_{s}\rightarrow \bar{\mu}\nu)}$ & $5.56\times10^{-3} $\\
\hline
$\nu_{i}\mu bu$   & $7.7\times10^{-5}$  & 
$BR(B^{+}\rightarrow \bar{\mu}\nu)$ & $<1.0\times10^{-6}$\\
\hline
$\nu_{i} \mu bc $  &     &  & \\
\hline
$\nu_{i} \tau du $  &  $8.0\times10^{-2}$     &$BR(\tau \rightarrow \pi^+ \nu)$  & $ (1.083 \pm 0.006) \times 10^{-1} $\\
\hline
$\nu_{i} \tau dc $  &  $5.2\times10^{-2}$   & $BR(D^{+}\rightarrow \bar{\tau}
\nu) $ & $<1.2 \times10^{-3}$\\
\hline
$\nu_{i} \tau su $ & $2.6\times10^{-2}$ & $BR(\tau \rightarrow K^+ \nu)$      
& $(7.0\pm 0.1) \times 10^{-3} $ \\
\hline
$\nu_{i} \tau sc$  & 
 $4.9\times10^{-2}$   
& $BR(D^{+}_{s}\rightarrow\bar{\tau}\nu )$ 
& $5.54 \times10^{-2} $\\
\hline
$\nu_{i} \tau bu $& $7.2\times 10^{-4}$  & $BR(B^{+}\rightarrow \bar{\tau}\nu_{\tau})$ & $(1.1 \pm 0.3) \times10^{-4}$\\ 
\hline
$\nu_{i} \tau bc $  &     &  & \\
\hline
\end{tabular}
\caption{Constraints  from ``charged current'' processes on
$S \pm P$ operators.
This bound applies to  
 $\epsilon^{ij\beta q}_{CC}$, in an  basis
using  tensor operators.
The first column is the index combination
$ij\beta q$,  the second is the constraints, which
 arise from the observable  given  in column 3.
The  experimental value  used is the last column.
$\nu_i$ is any flavour of neutrino. }
\label{tab:SPnu}
\end{center}
\end{table}

\subsection{Estimating the loops  }
\label{ssec:loops}

Consider first the $V \pm A$  operators 
${\cal O}^{YR}$, involving
the singlet $t_R$,
with coefficient $ 2\sqrt{2} G_F \epsilon_{YR}^{ijqt}$,
$q \in \{ c,u \}$.
 Dressing
with a $W$ loop as in 
Figure~\ref{fig:feyn}   gives
the operators
 $ {\cal O}^{YL}_{ij \alpha \beta} =$
 $(\overline{\ell}_i \g_\a P_Y \ell_j)$ 
$(\overline{d}_\alpha \g^\a P_L d_\beta)$ with 
 coefficients 
\bea
\epsilon_{YL}^{ij \alpha \beta}
\simeq -\frac{g^2 m_t m_{q} V_{t\beta} V^*_{q \alpha}}{16\pi^2  (m_t^2 - m_W^2) } 
\log \left( \frac{m_t^2}{m_W^2} \right)
\epsilon_{YR}^{ijqt}
\simeq -\frac{2\alpha_{em}}{3}
 V_{t\beta} V^*_{q\alpha}  \frac{m_{q}}{m_t}  
\epsilon_{YR}^{ijqt}
\label{V-AR}
\eea 
where $(\beta,\alpha) \in \{ (b,s), (b,d),(s,d) \}$, 
$q \in \{ c,u \}$.
 The log is $\simeq 1.54 \simeq \pi/2$.
The numerical factor 
arising from the loop
depends on the  light external quark  indices $\beta,\alpha$,
as well as the quark flavour $q$ participating 
with the top in the contact interaction.  It is given in
 Table~\ref{tab:nrm}. It is clear that the loop
gives significant suppression, due to CKM elements
and light quark masses.

The $V \pm A$  operators 
${\cal O}^{YL}$, which involve  doublet top quarks,
have a larger mixing to down-type quark operators.
 Dressing with a $W$ loop as in 
Figure~\ref{fig:feyn}   gives
the operator
 $
{\cal O}^{YL}_{ij\alpha\beta} =$
 $(\overline{\ell}_i \g_\a P_Y \ell_j)$ 
$(\overline{d}_\alpha \g^\a P_L d_\beta)$,
with  a log divergent
 coefficient,  unsuppressed by light quark masses:
\bea
 \epsilon_{YL}^{ij\alpha\beta}
\simeq  \frac{g^2  V_{t\beta} V^*_{q\alpha}}{32\pi^2 } 
 \left(\frac{m_W^2}{m_t^2 - m_W^2}
\log \left[ \frac{m_t^2}{m_W^2} \right]
-\log \left[ \frac{\Lambda^2}{m_t^2} \right]
 \right)
\epsilon_{YL}^{ijqt}
\simeq  - \frac{3\alpha_{em}}{4\pi}
 V_{t\beta} V^*_{q\alpha}    
\epsilon_{YL}^{ijqt}
\label{V-AL}
\eea 
where $(\beta,\alpha) \in \{ (b,s), (b,d),(s,d) \}$, 
$q \in \{ c,u \}$,  only terms with
logs were retained, and 
  $\Lambda \simeq 3 m_t$
in the log to obtain 
the second approximation\footnote{We will obtain a
bound  of $\epsilon \lsim 0.01$, which corresponds
to $\Lambda\gsim 10m_t$. We nonetheless
conservatively take $\Lambda \sim 3m_t$ to
allow for  New Particle masses $m_{NP}$ that are lower
than $\Lambda \propto m_{NP}/g_{NP}$.} and the 
numerical factors in Table~\ref{tab:nrm}.
 The  top operators ${\cal O}_{ijqt}^{LL}$
can also generate Charged Current (CC) operators
 $(\overline{\nu}_i \g^\rho P_L \ell_j)$ 
$(\overline{u}_q  \g^\rho P_L d_\beta)$,
suppressed only by $V_{t\beta}$, however the
bounds on the coefficients of such CC
operators are weaker than the bounds on
 FCNC operators, so the best limit on
the top operators arises from the
loop given in Eqs.~(\ref{V-AR},\ref{V-AL}).

The $S \pm P$ operators   ${\cal O}^{ S\pm P,X}_{ijqt}$
 involving either $t_L$ ($X=L$) or
$t_R$ ($X=R$),  
will generate $S \pm P$ operators
 for down-type quarks
 $ {\cal O}^{S\pm P,R}_{ij\alpha\beta} =$
 $(\overline{e}_i  P_Y e_j)$ 
$(\overline{d}_\alpha  P_R d_\beta)$,
where it is always the heavier $d_\beta$
that is right-handed, because there is
a mass insertion on an  external quark  leg.
The lepton current is unaffacted by the
loop, so will have the same chirality in the
induced down-quark operator, as it had in
the top-LFV operator. 
There is also a mass insertion on the right-handed
 internal quark line, so the coefficient
is $\propto m_t$ in the presence of   ${\cal O}^{ S\pm P,R}_{ijqt}$:
\bea
 \epsilon_{ S\pm P,R}^{ij \alpha \beta}
&\sim& 
\frac{g^2 m_t m_\beta V_{t\beta} V^*_{q\alpha}}{32\pi^2 (m_t^2-m_W^2)} 
\left( 1 - \frac{m_W^2}{m_t^2 - m_W^2}  \log \frac{m_t^2}{m_W^2}\right) 
\epsilon_{ S\pm P,R}^{ijqt}
\nonumber\\
&\simeq&
 \frac{\alpha_{em}}{3\pi }   V_{t\beta} V^*_{q\alpha}\frac{ m_\beta}{ m_t}   
\epsilon_{ S\pm P,R}^{ijqt} ~~. 
\label{S+P}
\eea
The result in the presence of
  ${\cal O}^{ LQ,R}_{ijqt}$ is similar: the $LQ$  operator
 for down-type quarks,  $ {\cal O}^{LQ,R}_{ij\alpha\beta} =$
 $(\overline{d}_\alpha  P_R e_j)$ 
$(\overline{e}_i  P_R d_\beta)$,
is generated with a coefficient 
obtained by replacing  $\epsilon_{ S\pm P,R}^{ijqt} 
\to -\epsilon_{ LQ,R}^{ijqt}$ in Eq.~(\ref{S+P}).

In the presence of ${\cal O}^{ S\pm P,L}_{ijqt}$ 
(or ${\cal O}^{ LQ,L}_{ijqt}$),
the loop factor  is  $\propto m_q m_\beta$,
which is negligibly small. So 
we obtain no interesting bounds on the
 ${\cal O}^{ S- P,L}_{ijqt}$ operators.
 However,   the  operators
${\cal O}^{S+P,L}_{ijqt}$ (and $ {\cal O}^{LQ,L}_{ijqt}$)
contain $t_L$ and $e_{Lj}$,
 between whom the $W$ could  be exchanged 
 as indicated in the
right diagram of Figure~\ref{fig:feyn}.
This dressing of
the top-operator ${\cal O}^{S+P,L}_{ijqt}$
will generate  ${\cal O}^{CC}_{ijqt \b}$ and
a tensor operator, in the linear
combination corresponding to
${\cal O}^{CC,LQ}$ 
of Equation~(\ref{CC}).
The loop coefficient is unsuppressed
by masses,  as in Equation~(\ref{V-AL}):
\bea
 \epsilon_{CC,LQ}^{ij q\beta}
&\simeq&  -\frac{ g^2  V_{t\beta} }{16 \pi^2 } 
\left(\frac{m_W^2}{m_t^2 - m_W^2}
\log \left[ \frac{m_t^2}{m_W^2} \right]
-\log \left[ \frac{\Lambda^2}{m_t^2} \right]
 \right)
\epsilon_{S+P,L}^{ijqt}
\simeq   \frac{2\alpha_{em}}{\pi}
 V_{t\beta}     
\epsilon_{S+P,L}^{ijqt} ~~~, \nonumber\\
 \epsilon_{CC}^{ij q\beta}
&\simeq  & \frac{2\alpha_{em}}{\pi}
 V_{t\beta}     
\epsilon_{LQ,L}^{ijqt} ~~~.
\label{funny1}
\eea 
Or, in a basis with tensor operators
\bea
 \epsilon_{CC}^{ij q\beta}
&\simeq&  \frac{ g^2  V_{t\beta} }{32 \pi^2 } 
\left(\frac{m_W^2}{m_t^2 - m_W^2}
\log \left[ \frac{m_t^2}{m_W^2} \right]
-\log \left[ \frac{\Lambda^2}{m_t^2} \right]
 \right)
\epsilon_{S+P,L}^{ijqt}
\simeq  - \frac{\alpha_{em}}{\pi}
 V_{t\beta}     
\epsilon_{S+P,L}^{ijqt} ~~~. 
\label{funny}
\eea

\begin{table}[htb]
\begin{center}
\begin{tabular}{||  c | c  |c|c||   }
\hline \hline
top op. & down coeff.  &  $q=u$& $q=c$\\
\hline
&$\epsilon_{XL}^{ij  db}$&$ 8.0 \times  10^{-8 }$&  $7.7 \times  10^{-6}$\\
${\cal O}^{XR}_{ijqt}$ 
&$\epsilon_{XL}^{ijsb}$&$1.8  \times  10^{-8 }$&  $3.4 \times  10^{-5}$\\
&$\epsilon_{XL}^{ij ds}$&$3.2  \times  10^{-9 }$&  $3.1 \times  10^{-7}$\\
\hline
&$\epsilon_{XL}^{ijdb}$&$ 2.0 \times  10^{- 3}$&  $4.6 \times  10^{-4}$\\
 ${\cal O}^{XL}_{ijqt}$
&$\epsilon_{XL}^{ij sb}$&$ 4.6 \times  10^{-4 }$&  $ 2.0 \times  10^{-3}$\\
&$\epsilon_{XL}^{ij  ds}$&$ 8.2 \times  10^{-5 }$&  $1.9 \times  10^{-5}$\\
\hline
&$\epsilon_{S \pm P,R}^{ij   db}$ 
&$ 2.4 \times  10^{-5 }$&  $5.3 \times  10^{-6}$\\
 ${\cal O}^{S\pm P,R}_{ijqt}$ 
&$\epsilon_{S \pm P,R}^{ij   sb}$
&$5.3  \times  10^{- 6}$&  $2.4 \times  10^{-5}$\\
&$\epsilon_{S \pm P,R}^{ij   ds}$
&$4.1  \times  10^{-9 }$&  $1.8 \times  10^{-8}$\\
\hline
&
$\epsilon_{CC}^{ij   qb}$  
&$  2.5\times  10^{- 3}$&  $ 2.5\times  10^{-3}$\\
 ${\cal O}^{S+ P,L}_{ijqt}$ 
&
$\epsilon_{CC}^{ij   qs}$ &$ 1.0\times  10^{- 4}$&  $1.0 \times  10^{-4}$\\
& 
$\epsilon_{CC}^{ij   qd}$&$ 2.0 \times  10^{-5 }$&  $ 2.0\times  10^{-5}$\\
\hline
 \hline
\end{tabular}
\end{center}
\caption{ The last two columns
give the numerical value of the coefficient 
given in the second column,  generated by
loop containing the  top  LFV  operator given in the first
column, of coefficient $2 \sqrt{2} G_F$.
The top LFV operator involves the top and 
and  the quark $u_q\in \{u,c \}$; the two choices of
$q$ identify the last two columns.
The operators ${\cal O}^{S\pm P,L}_{ijqt}$ 
also induce
FCNC operators (involving two down-type
quarks), but the coefficients  are not listed because
they are suppressed by two light
quark masses.
\label{tab:nrm}}
\end{table}

\subsection{Bounds and cancellations}
\label{ssec:bounds}

The numbers in  
   Table~\ref{tab:nrm} can be compared to
the bounds  on the various
$\epsilon$s for light quark operators,
 given  in the Tables~\ref{tab:eedd}, \ref{tab:SP}
and \ref{tab:SPnu}. For instance,
the last column of the third line   of  Table~\ref{tab:nrm} says that
 $\epsilon^{ijct} _{XR}\frac{4G_F}{\sqrt{2}} 
(\overline{\ell}_i \g_\sigma P_X \ell_j)
(\overline{c} \g^\sigma P_R t)$
generates 
 $\epsilon_{XL}^{ijds}\frac{4G_F}{\sqrt{2}}
(\overline{\ell}_i \g_\sigma P_X \ell_j)$$ (\overline{d} \g^\sigma P_L s)$ 
with  $$\epsilon _{XL}^{ijds} = 3.1 \times  10^{-7} \epsilon^{ijct} _{XR}
\lsim  3.0 \times  10^{-7}
~~~\Rightarrow~~~\epsilon^{e \mu ct} _{XR} \lsim 1
$$
where  $\epsilon^{e\mu ds} _{YL} \lsim  3.0 \times  10^{-7}$  
 is the upper
bound from $K\to e\bar{\mu}$ given in  Table~\ref{tab:eedd}, which  
sets the  constraint  $\epsilon^{e \mu ct} _{XR} \lsim 1$.
    This bound on the $e \mu ct$  coefficient
of ${\cal O}^{XR}$ operators
appears in  the second row
and  last two columns
of Table~\ref{tab:topVA}.  In the case
of $e\tau c t$ and $\mu \tau c t$  indices, 
the best bound  on  the coefficient of the 
 ${\cal O}^{XR}$ operators arises from
$B$ decays. This  can be seen because Table~\ref{tab:topVA} 
gives the origin of the bound as
${\cal O}^{LL}_{i\tau db}$, and the bound on
these coefficients from Table~\ref{tab:eedd}
is from $B$ decays\footnote{ We obtain 
$\epsilon^{ \mu \tau ut}_{LR}$,
$\epsilon^{ \mu \tau ut}_{RR}
<2 \times 10^5$
from the limit on $\epsilon^{ \mu \tau db}_{LL}$,
but do not include thes numbers in
Table~\ref{tab:topVA} because they are to 
weak to be meaningful.}.

The bounds of Table~\ref{tab:topVA} implicitly assume
that only one LFV top operator is present
at a time.  This assumption depends on
the choice of operator basis, so it is interesting
to  consider the possibility
of cancellations among the $V \pm A$ operators. 
Many of the bounds arise from pseudoscalar
down-type meson decays, which are mediated
by an operator of the form
$(\overline{e}_i \g^\rho P_X e_j )(\overline{d}_\a \g^\rho \g_5 d_\beta )$,
in which case  the bound applies to  the combinations
$\epsilon^{ij \a \b} _{XR} - \epsilon^{ij \a \b} _{XL}$. 
Since the  loop
that transforms  top-operators into
down-type quark operators  gives different
suppression factors for operators
involving $t_L$ or $t_R$, we neglect
possible cancellations among $XL$ and $XR$  operators. 
So we neglect the possbility of cancellations
for $V\pm A$ operators.

{
In obtaining bounds from 
(pseudoscalar) meson decays, an operator basis
that includes tensors is convenient,  
because the tensors  do not contribute~\cite{notensor} to
the decays  $M \to \ell_i \bar{\ell}_j$, where $M$ 
is the pseudoscalar
meson made of $q_1\bar{q}_2$ and
$\ell_i $ are  neutrinos or charged leptons. 
So we quote low-energy bounds in a basis with
tensors\footnote{Alternatively,
one can impose pseudoscalar decay bounds  on
$\epsilon^{S+P,X}_{i j q_2 q_1}  -\frac{1}{2}\epsilon^{LQ,X}_{i j q_2 q_1}$.
From Eq.~(\ref{S+P}) and after, the top-operators
${\cal O}^{S +P,R}$ and ${\cal O}^{LQ,R}$ induce
respectively  $S+P$ and $LQ$ operators  
involving down-type quarks.
So in Table~\ref{tab:topSP} are quoted bounds on
$\epsilon^{S+P,R}_{i j qt}  -\frac{1}{2}\epsilon^{LQ,R}_{i j qt}$.
The $t_L$ operators
${\cal O}^{S +P,L}$ and ${\cal O}^{LQ,L}$ 
respectively induce the  ${\cal O}^{CC} $  and  ${\cal O}^{CC,LQ}$ 
operators (see Eq.~(\ref{funny1})), which mediate
 the leptonic decays of charged mesons, such
as $B^+ \to \nu e^+$. So in this case
the bound applies to
$\epsilon^{LQ,L}_{i j qt}  -\frac{1}{2}\epsilon^{S+P,L}_{i j qt}$
as quoted in Table~\ref{tab:topSP}.}.

}

\section{Appendix:  single $t$  production at HERA}
\label{ssec:hera}

HERA was an $e^\pm p$ collider with $\sqrt{s} = 319$ GeV,
where the H1~\cite{H109} and ZEUS~\cite{ZEUS} experiments searched for single
top production: $\sigma (e^\pm p \to e^\pm t X)$. We follow an H1
analysis~\cite{H109}, which is outlined in Section
\ref{ssec:H1}. Then in Section~\ref{ssec:CI@hera},
we discuss how to translate the results of
this analysis  to a bound on
the LFV process
$\sigma (e^\pm p \to \mu^\pm t X)$, and in
Section~\ref{ssec:sigmas}, we estimate 
$\sigma (e^\pm p \to \mu^\pm t X)$
in the presence of the operators
of Appendix~\ref{ssec:CIs}.

\subsection{The H1 analysis}
\label{ssec:H1}
The H1 collaboration set a bound~\cite{H109}: 
\beq
\sigma (e^\pm  p \to e^\pm t +X) \leq 0.30  ~~ {\rm pb}
 =
\frac{2.3\times 10^{-5}}{m_t^2}
~~{\rm  at}~ 95\% CL.
\label{H1bd_bis}
\eeq
using  474 pb$^{-1}$ of data, and requiring
the top decay $t \to \mu^+ \nu b$ \footnote{H1 obtained
 a more restrictive bound, $\sigma \leq 0.25$ pb, by combining
the various top decay modes}.  H1 does not give
the luminosity  in $e^+p$ and $e^-p$, but from~\cite{H103,1406}, we   interpret  
$\geq$ 228 pb$^{-1}$ of  $e^+p$ data and
$\geq$ 205 pb$^{-1}$ of  $e^-p$ data
 at $\sqrt{s} = 319$ GeV, so we approximate
that half the luminosity was in  $e^+p$,
and half in   $e^-p$.

We would like to use this limit to set a bound on
$
\sigma (e^\pm  p \to \mu^\pm t +X) 
$. This is possible because H1 searched for leptonic
decays of the top, $t\to \mu^+ \nu b$, and did not
require to observe the  $e^\pm$. As a first
step in a cut-based analysis, 
H1 required isolated muons with $p_T> 10 $ GeV,
and missing energy  $\ETm >12$ GeV; they  found
 14  events where they expected $11 \pm 1.8$ Standard Model background.
Their signal efficiency, for a magnetic moment
$ut\g$ coupling, was 44\%.
A $e^\pm u \to \mu^\pm t$ interaction mediated
by our top operators should pass this first set of cuts,
so we would like to obtain a limit on
the coefficients of our operators from this stage
of the analysis. 
  Then H1 required
the muon to be positively charged, and
in combination  with the jet and $\ETm$, 
to reconstruct to  a top.
Only  four events remain after imposing these
cuts,
with $2.1  \pm 0.3$ expected in the Standard Model. 
H1's signal efficiency was 36\% and at this point
they obtain the limit 
$\sigma (e^+ p\to (\mu^+ \ETm  b)  e^+  X) \leq 0.30$ pb.

\subsection{Translating to contact interactions}
\label{ssec:CI@hera}

To translate the H1 bound to our contact interactions,
we must address various issues.
\ben
\item We estimate that the  bound on
$\sigma (e^+ p\to (\mu^+ \ETm  b)  e^+  X)$
after the first cut, is comparable to the final  bound 
obtained by H1, so we use
the limit of Eq.~(\ref{H1bd_bis}). To obtain
this curious approximation, 
 we estimate the
$t$ production cross section that could be excluded
at 95\% C.L.  using Poisson statistics for
4 observed events with $2$ background expected
\footnote{That is, the number of signal events
$s$ that give 0.05 = $P(s+b)/P(b)$ where
$b=$ background expected.}, and then
we estimate the
$t$ production cross section that  would
give a gaussian distribution
of events centered  $4\sigma$ above 11, 
that is at 18. And we find these cross sections
are comparable.

\item  Different operators contribute to 
$e^+ p\to \mu^+  t+ X$ and 
$e^- p\to \mu^-  t+ X$. Since the
luminosity is approximately equally
distributed between $e^+ p$ and 
$e^- p$, we suppose that the
bound for an individual operator is
$\sigma \lsim 2\times 0.30 $ pb. 
\item In the case of our operators,
a central $\mu^\pm$ is produced by
the contact interaction, so the
top can decay to any lepton (we do not allow
for hadronic $t$ decays, because they
are unlikely to produce $\ETm \geq 10$ GeV).
This means the upper bound in the cross section
is reduced by 1/3, because H1 required
that the top decay to $\mu^+  \nu b$. 
\item We need an estimate for the efficiency
for our operators. H1 gives a signal efficiency
of 44\% after their first cut, in part
because they detect muons as a $p_T$ imbalance
in the calorimeter, with a 60\% efficiency
for detecting a muon of $p_T$ = 10 GeV.
Since our event has a central muon, like
H1's signal, we take the efficiencies
to be the same, and assume this makes
our bounds uncertain by a factor $\sim 2$. 
\een
So in summary, we  multiply the H1 bound
by 2/3, and impose
\beq
\sigma (e^\pm p\to \mu^\pm  t  +X) \leq 0.20  ~~ {\rm pb}
 = \frac{1.5 \times 10^{-5}}{m_t^2}
~~{\rm  at}~ 95\% CL.
\label{ourbd}
\eeq

\subsection{$\bm{\sigma (e^\pm p \to \mu^\pm t +X)}$ in the presence of top operators}
\label{ssec:sigmas}

The operators
of Section~\ref{ssec:CIs}
induce
a differential partonic  cross section for
$e^\pm u \to \mu^\pm t$ that does 
  not diverge for emission at zero angle. 
We therefore  integrate the
partonic cross section
 over all angles (neglecting experimental cuts), to obtain:
\bea
\hat{\sigma} (e^\pm u \to \mu^\pm t)
&=& |\epsilon_{XX}|^2  \frac{( 1 -m_t^2/\hats)^2}{16 \pi m_t^4} 
\hats
 ~~~~~~~~~~~~~~~~~~~~~~~~~~  {\cal O}_{LL}~,~  {\cal O}_{RR} \label{sigV+A}\\
&=& |\epsilon_{YX}|^2  \frac{( 1 -m_t^2/\hats)^2}{48 \pi m_t^4}
(\hats + \frac{m_t^2}{2})
~~~~~~~~~~~~~~~~~~~~~~  {\cal O}_{LR}~,~  {\cal O}_{RL} \label{sigV-A}\\
&=& |\epsilon_{S-P,X}|^2  \frac{( 1 -m_t^2/\hats)^2}{192 \pi m_t^4}
(\hats + \frac{m_t^2}{2})
~~~~~~~~~~~~~~~~~~~~~~~~  {\cal O}_{S-P,X} \nonumber\\
&=& 
\frac{( 1 -m_t^2/\hats)^2}{192 \pi m_t^4}  
\left(
[
|\epsilon_{S+P,X}   +  \epsilon_{LQ,X}|^2]
[\hats + \frac{m_t^2}{2}] 
+  {\rm Re}\{\epsilon_{S+P,X} \epsilon_{LQ,X}\} [\hats  -m_t^2]
 \right) 
 ~~~~~~{\cal O}_{S- P,X}~ ,~ {\cal O}_{LQ,X}  
  \nonumber
\eea

To obtain the cross section on the proton, 
we define $x =\hats/s$, and obtain,
 integrating over CTEQ10~\cite{CTEQ10}  parton distribution functions:
\bea
\int_{0.33}^{1}  dx  x (1-.295/x)^2  f_u(x)  = 0.0046 
&&,~~ \int_{0.33}^{1}  dx  x  f_u(x)  = 0.041 \nonumber\\ 
~~~\int_{0.33}^{1}  dx  (1-.295/x)^2  f_u(x)  = 0.0092
&&,~~  \int_{0.33}^{1}  dx   f_u(x)  = 0.096
\label{ubi1}
\eea
where the lower  $x$ integration limit
is estimated  to produce,  in the centre-of-mass frame,
  a top of negligible
momentum, and a muon of $p_T > 10$ GeV.
The $(1-.295/x)^2$ represents the kinematic
suppression factor $(1-m_t^2/\hats)^2$.

The resulting cross sections   are 
(for $s/m_t^2 = 3.5$): 
\bea
\sigma (e^\pm p \to \mu^\pm t+X)
&=& 
0.0046 s  \frac{|\epsilon_{XX}|^2}{16 \pi m_t^4} 
 ~~~ \Rightarrow 
\epsilon_{XX}^{e\mu u t}, \epsilon_{XX}^{\mu e u t} < 0.22
 \label{sigfV+A}\\
&=& 
 \frac{|\epsilon_{XY}|^2}{48 \pi m_t^4}
0.0046( s +m_t^2)
 ~~~ \Rightarrow 
\epsilon_{XY}^{e\mu u t}, \epsilon_{XY}^{\mu e u t} < 0.33
\label{sigfV-A}\\
&=& 
 \frac{|\epsilon_{S-P,X}|^2}{192\pi m_t^4}
0.0046( s +m_t^2)
 ~~~ \Rightarrow 
\epsilon_{S - P ,X}^{e\mu u t} ,
\epsilon_{S - P ,X}^{\mu e u t}  < 0.66
 \\
&=& 
 \frac{|\epsilon_{S+P,X}|^2 + |\epsilon_{LQ,X}|^2}{192\pi m_t^4}
0.0046( s +m_t^2)
 ~~~ \Rightarrow 
\epsilon_{S + P ,X}^{e\mu u t} ~,~ 
\epsilon_{S + P ,X}^{\mu e u t} ~,~ 
\epsilon_{LQ ,X}^{e\mu u t} ~,~
\epsilon_{LQ ,X}^{\mu e u t}  < 0.66
 \nonumber
\eea
where $X,Y \in \{L,R\}$ and  the bounds are estimated  by
imposing
the  bound  of Eq.~(\ref{ourbd}).

\subsection{What about $\epsilon^{e\tau ut}$ at HERA?}
\label{app:tilde}

The $\epsilon^{e\tau ut}$, $\epsilon^{\tau e ut}$ 
are poorly constrained by rare decays,
whereas $e^\pm u \to \tau^\pm t$ could have occured at
HERA. It is unclear to the authors how H1 would
have treated a hadronic $\tau$ decay,
so we  conservatively restrict to  $\tau \to e\nu \bar{\nu}$ decays.
Allowing any decay of the  top, we use the combined
H1 bound 
\beq
\sigma (e^+p\to e^+ t +X) \leq 0.25  ~~ {\rm pb}
= \frac{1.9\times 10^{-5}}{m_t^2}
~~{\rm  at}~ 95\% CL,
\label{H1combobd}
\eeq
multiplied by $2/BR(\tau \to e\nu \bar{\nu})  \simeq$ 12 ,
where the 2 is from point 2 of Section
\ref{ssec:CI@hera} above. 
 So
$\sigma (e^\pm p\to \tau^\pm t +X) \lsim 3.0  ~~ {\rm pb}$,
and the approximate bounds on 
$\epsilon^{e\tau u t}$, $\epsilon^{\tau e u t}$
are given in Table~\ref{tab:topVA}.


\section{Appendix:  The CMS search for $t\longrightarrow Zq$}
\label{app:CMStZq}
The CMS collaboration sets the limit $BR(t\to Z q) \leq 6\cdot 10^{-4}$~\cite{CMS:tZc}, by searching for $t\bar{t}$ production, 
with one  top decaying leptonically, and the other decaying to 
$Zq$ followed by $Z\to \ell\bar{\ell}$.
In this Section, we emulate in our framework 
this published CMS analysis, which
searches for a  final state similar to ours. 

We hence implement in our framework the selection described in~\cite{CMS:tZc}. 
The definition of good jets and isolated leptons is the same as in our 
analysis  described in Section~\ref{sec:sel}. 
We select events containing exactly 3 isolated 
charged leptons (electrons or muons), 
 of which two have the same flavor and opposite sign. 
Events are requested to contain at least 2 jets, 
and exactly one b-tagged jet, and the missing 
transverse energy has to be higher than 30 GeV. 
The invariant mass of the opposite sign and 
same flavor lepton is required to lie 
within 78 and 102 GeV/c$^{2}$.

The two opposite sign and same flavor leptons are identified as coming from the Z decay, if there is 
more than one possible 
lepton pair, the one with
invariant mass closest to the Z mass ($m_{Z}$)  
is chosen. 
The remaining lepton is associated with 
the decay of the W and used, with the missing transverse energy, 
to calculate the neutrino longitudinal 
momentum with a W mass constraint, 
as explained in Section~\ref{sec:sel}. 
The invariant mass of the W and the b-tagged jet is required to 
be within 35 GeV/c$^{2}$ of the nominal top mass.  
The 4-momenta of the two leptons coming from the Z 
are combined with all (non b-tagged) jets in the event. 
In order for the event to be selected, at least one of 
these combinations must have a resulting invariant 
mass within 25 GeV/c$^2$ of the nominal top mass. 

All the background processes taken into account in the CMS analysis are considered. 
In our simulated samples, the expected  number of background events
after applying  the CMS selection is $1.5\pm 0.2$, 
to compare with the expectation from simulation of 
the public CMS analysis, of $3.2\pm 1.2 \mathrm{(stat)} \pm 1.5\mathrm{(syst)}$. 
This comparison is not meant to be 
rigorous, as the Delphes emulation of the CMS detector reconstruction is known to be imperfect,
and the samples used here do not include any effect from pileup. Nevertheless, the agreement is good enough to validate our framework as 
a tool to extract reasonable sensitivity studies. The result of this comparison also motivates the variation of 100\% 
in the expected background events for the limit calculation, performed in the main study (Section \ref{sec:limits}).

The limit setting procedure is validated as well: when using the number of expected events from the public result in our statistical procedure, we obtain an expected upper limit of $BR(t\to Z q)<0.9\cdot 10^{-3}$ to compare to the public result $BR(t\to Z q)<1\cdot 10^{-3}$. 

Having validated our method on expected background 
allows us to evaluate the constraint set  on the $\topqme$ 
branching ratio by the CMS analysis~\cite{CMS:tZc}.
Indeed, our LFV top events and the CMS
analysis share  the same overall final state (missing transverse energy, 
3 isolated leptons, and 2 jets, of which one is a $b$-jet). Half of
our LFV events should give a pair of opposite sign,
same flavour leptons, as required  in the CMS
analysis. However, in our case, they do not come
from the $Z$, but respectively from the $t$ and
$\bar{t}$ (see Figure~\ref{fig:diagram_signal}), so 
they are easily rejected by the CMS requirement that their invariant
mass be near $m_Z$.
In order to quantify this, we evaluate the efficiency of  the CMS event selection on  the $\topqme$ signal sample, obtaining $\epsilon=0.050\pm0.005 \%$. 
This would correspond to an expected limit on LFV in top decays of $BR(\topqme)<3.7\cdot 10^{-3}$, according to the limit setting procedure based on RooStats (a simple rescaling of the published limit with selection efficiencies and cross sections would give a comparable value of $BR(\topqme)<4\cdot 10^{-3}$).  
Comparing to  Equation~(\ref{BRtop}), we see that
this is on the verge of probing LFV top decays mediated by a
four-fermion operator.

\bibliography{d2}
\bibliographystyle{plain}

\end{document}
\bye